\documentstyle[amssymb,epsfig,aps]{revtex}

\begin{document}
\title{Semiclassical catastrophes and accumulative angular squeezing of a 
kicked quantum rotor}
\author{M. Leibscher$^{1}$, I.Sh.Averbukh$^{1}$, P. Rozmej$^{2}$, and R. Arvieu$^{3}$}
\address{$^{1}$Department of Chemical Physics, The Weizmann Institute of Science,\\
Rehovot 76100, Israel\\
$^{2}$Institute of Physics, University of Zielona G\'ora, 65246 Zielona
G\'ora,\\
Poland\\
$^3$Laboratoire de Physique Subatomique et de Cosmologie \footnote{formerly: 
Institut des Sciences Nucl\'eaires}, 38026 Grenoble Cedex, France}
\maketitle

\begin{abstract}
We present a detailed theory of spectacular semiclassical catastrophes happening
during the time evolution of a kicked quantum rotor 
(Phys.Rev. Lett. {\bf 87}, 163601 (2001)).
Both two- and three-dimensional rotational systems are analyzed. It is shown
that the wave function of the rotor develops a {\em cusp} at a certain delay after a kick,
which results in a sharply focused rotational wave packet. The {\em cusp} is followed by a fold-type
catastrophe manifested in the {\em rainbow}-like moving angular singularities. In the 
three-dimensional case, the rainbows are accompanied by additional singular features similar
to {\em glory} structures known in wave optics. These catastrophes in the time-dependent 
angular wave function are well described by the appropriate tools of the quasiclassical
wave mechanics, i.e. by Airy and Bessel approximations and Pearcey's functions. A scenario
of "accumulative squeezing" is also presented in which a specially designed train of short
kicks produces an unlimited narrowing of the rotor angular distribution. This
scenario is relevant for the molecular alignment by short laser pulses, and
also for atom lithography schemes in which cold atoms are focused by an optical
standing wave.

PACS numbers: 32.80.Lg, 03.65.Sq, 05.45.-a
\end{abstract}

\section{Introduction}

The driven rotor is a standard model for classical and quantum-mechanical
nonlinear studies. The interest to the subject has substantially increased
because of \ new laser experiments on molecular alignment (orientation)
(for a recent review see \cite{Sta03a}), and
due to the quantum optics realization of the $\delta $-kicked rotor \cite
{Moore,Ammann}. Alignment of molecules by using laser pulses that are much
shorter than the typical molecular rotational times is of special interest.
At these conditions, a coherent rotational wave packet is generated that
takes an angularly squeezed shape after the pulse is over \cite
{lin75,fon88,Bandrauk,fel92,sei95,sei99,ort99,cai01,mac01,ros02,dio02,PRL2}.
This phenomenon opens new prospects for numerous applications in chemical
physics and non-linear optics which require enhanced molecular orientation
(alignment) for a limited time but at field-free conditions. Generation of
ultra-short laser pulses \cite{short,Ivanov} and control of high harmonics
generation \cite{harmonics} are only few examples to mention.

Motivated by these reasons two of us have revisited the problem of $\delta $%
-kicked quantum rotor recently \cite{rotor}. It was shown that free
evolution of a strongly kicked rotor is subject to several semiclassical
catastrophes with course of time, which are manifested in the angular
distribution singularities. In particular, a {\em cusp} develops in the
angular wave function at certain time after the pulse, providing an
extremely narrow peak in the direction of the kick. This effect is a
time-domain analog of the focusing phenomenon in wave optics. Moreover, the
subsequent spread of the focused rotational wave packet demonstrates a wave
front {\em fold }leading to the moving rainbow-type singularities in the
angular distribution. These effects can be observed both in two-dimensional
and three-dimensional rotational systems. The 2D rotational model is
important not only due to its relative simplicity, but also because it
describes the interaction of an ensemble of cold atoms with sinusoidal
standing light waves. In 3D case the geometry of the wave bifurcations is
even richer \cite{glory}. In particular, the angular cusp (focus) transforms
not only to the moving rainbow feature, but also to a persistent singular
peak in the direction of the kick. A little later such a peak appears also in
the opposite direction. These two peaks are reminiscent of the forward and
backward glories well known in the scattering of light by small droplets 
\cite{Hul81a}. As time goes on, the rotational wave packet exhibits a rich
scenario of fractional revivals \cite{Ave89a} with the presence of many
angular rainbows and foci.

The cusp, fold and glory are well known catastrophes in wave optics and have
been studied since long ago (see, e.g. \cite{Tho75a,Berry} and references
therein). A complete study of the morphologies of caustics in optics has been
given in \cite{Ber80a}. Catastrophes are also known in the field of atomic
and nuclear physics where some of them (like rainbow and glory) were studied
at a very early stage \cite{Ford}. Besides many experimental works to unravel
caustics \cite{eis61,bry68}, a large number of theoretical studies were
devoted to the semiclassical description of the occurring
singularities \cite{ber72,bri85,ada02,Ber80a,Nus92a}. The simplest
semiclassical approaches are based on the stationary phase approximation for
integral representations of the related wave functions. They can be
successfully used if the stationary phase points are well separated from
each other. However, the semiclassical catastrophes like rainbow and glory
involve different kinds of non-isolated and merging stationary points.
Various uniform approximations have been developed for such a case \cite
{Ber66a,ber69,con73,con81}.

In the present study we apply semiclassical methods to the time-domain cusp,
fold and glory phenomena exhibited by rotors subject to strong kicks. At the
beginning of Section II we specify two different kinds of coupling to the
external field (dipole-type and polarization-type interaction). For
pedagogical reasons, we then continue with the classical analysis of the
angular distribution of an ensemble of kicked rotors. We start with the 2D
case, and consider additional effects specific to three-dimensional geometry
after that. The same approach is kept in the following parts of Section
II, where\ a fully quantum description of the problem is given. \ Section III
presents various integral representations \ for the wave function of a
kicked quantum rotational system. The next \ three Sections are devoted to
semiclassical theory of the time-domain focusing, rainbow and glory effects
for the kicked rotational systems, respectively . In Section VII we provide a
classical analysis of the time-dependent kicked rotor angular distribution
at finite temperature. Section VIII considers a systematic squeezing of the
rotor angular distribution by a train of multiple pulses. In \cite{rotor} we
proposed a multi-pulse excitation scheme that exhibits an accumulative
squeezing when increasing the number of kicks. After that, different
strategies leading to enhanced squeezing have been suggested, compared and
optimized for an ensemble of cold atoms in a pulsed optical lattice \cite
{Lei02} and molecular systems \cite{PRL2}. Recently, accumulative
squeezing was demonstrated experimentally in the atom optics system \cite
{Osk02a}. Section VIII provides details of the asymptotic behavior of the
accumulative squeezing scenario in the quasiclassical regime. Finally, the
results of the paper are summarized in Section IX. The paper has also two
Appendices with details of calculations.

\section{Kicked rotor system}

Under certain conditions, the process of molecular orientation (or
alignment) by laser fields can be described by a strongly driven 3D rigid
rotor model. The Hamiltonian of a 3D driven rotor is 
\begin{equation}
H=\frac{{\vec{L}}^{2}}{2I}+V(\theta ,t),  \label{3drotor}
\end{equation}
where $L$ is the angular momentum of the rotor, and $I$ is its moment of
inertia. For a linear molecule, having a permanent dipole moment $\mu $, and
driven by a linearly polarized field, the interaction potential is 
\begin{equation}
V(\theta ,t)=-\mu {\cal E}(t)\cos (\theta )  \label{permanent}
\end{equation}
where ${\cal E}(t)$ is the field amplitude (i.e., of a half-cycle pulse) and 
$\theta $ is the polar angle between the molecular axis and the field
direction. In the absence of interaction with a permanent dipole moment, the
external field couples with the induced molecular polarization. For
nonresonant laser fields, this interaction, averaged over fast optical
oscillations, is (see, e.g.,\cite{Boyd,Fri99}) 
\begin{equation}
V(\theta ,t)=-\frac{1}{4}{\cal E}^{2}(t)[(\alpha _{\Vert }-\alpha _{\bot
})\cos ^{2}(\theta )+\alpha _{\bot }].  \label{polarization}
\end{equation}
Here $\alpha _{\Vert }$ and $\alpha _{\bot }$ are the components of the
polarizability, parallel and perpendicular to the molecular axis, and ${\cal %
E}(t)$ is the {\em envelope} of the laser pulse. The quantized motion of a
rigid rotor is convenient to describe using dimensionless time scale, $\tau
=t\hbar /I$. Introducing also the dimensionless interaction strength $%
\epsilon _{p}=\mu {\cal E}(t)I/\hbar ^{2}$ and $\epsilon _{i}=1/4{\cal E}%
^{2}(t)[(\alpha _{\Vert }-\alpha _{\bot })I/\hbar ^{2}$, the Hamiltonian Eq.(%
\ref{3drotor}) can be written either as 
\begin{equation}
H=\frac{{\vec{L}}^{2}}{2}-\epsilon _{p}(\tau )\cos \theta ,
\label{permdipole}
\end{equation}
or as 
\begin{equation}
H=\frac{{\vec{L}}^{2}}{2}-\epsilon _{i}(\tau )\cos ^{2}\theta ,
\label{inddipole}
\end{equation}
for the two above interaction types.

In this work, we consider laser pulses that are so short that the motion of
the rotor is ''frozen'' during the interaction with the pulse ($\delta $- kicked
rotor). In this case, the action of the laser field is characterized by the
time-integrated interaction intensity (kick strength) defined as 
\[
P=\int_{-\infty }^{\infty }\epsilon _{p}(\tau )d\tau 
\]

Although we started our discussion with a three-dimensional rotor, it is
equally interesting to study also the related two-dimensional systems. This
is justified both pedagogically and by the existence of physical systems
(i.e. cold atoms in standing optical waves \cite{Lei02,Osk02a}) that are described \ by such a
model.

\subsection{Classical description of an ensemble of 2D kicked rotors.}

We start with the classical dynamics of an ensemble of free 2D rotors \
having a permanent dipole moment (see Eq.(\ref{permdipole})), which are
kicked by a short pulse. The second interaction type (Eq.(\ref{inddipole}))
will be treated briefly after that.

Consider a collection of motionless rotors subject to a pulse applied at $%
\tau =0$. A particular rotor located initially at an angle $\theta _{0}$
will be found at the same angle just after the kick, but having the angular
velocity $-P\sin (\theta _{0}).$ The time-dependent position of this rotor
at time $\tau $ is given by \ \ \ \ \ \ 
\begin{equation}
\theta =\theta _{0}-P\tau \sin (\theta _{0})\text{ \ \ \ \ \ \ \ \ \ \ \ \ \
\ }(%
\mathop{\rm mod}%
2\pi ).  \label{mapping}
\end{equation}
For rotors starting their motion in the region of small angles ($\theta
_{0}<<1$), the acquired velocity is linearly proportional to the initial
angle. As the result, all such rotors will arrive at $\theta =0$ (''focal
point'') at the same time $\tau _{f}$: 
\[
\tau _{f}=1/P 
\]
providing a sharp peak in the angular distribution. This phenomenon is quite
similar to the focusing of light by a thin optical lens. Certainly, the
angular focusing is not perfect because of the aberration mechanism
(deviation of the $\cos (\theta )$ potential from the parabolic one). For $%
\tau <\tau _{f}$ ,\ Eq.(\ref{mapping}) presents a one-to-one mapping $\theta
(\theta _{0})$ (see Fig. \ref{clmap2} (a)). At $\tau =\tau _{f}$ the curve $%
\theta (\theta _{0})$ touches the horizontal axis (see Fig. \ref{clmap2}
(b)), i.e. the mapping turns degenerate. At $\tau >\tau _{f}$ the angle $%
\theta _{0}$ becomes a multi-valued function of $\theta $ (Figs. \ref{clmap2}
(c) and \ref{clmap2} (d)). The probability of finding a rotor in a certain
angle element $d\theta $ is determined by the initial distribution function, 
$f_{0}(\theta _{0})$ as follows: 
\begin{equation}
f(\theta )d\theta =\sum_{a}f_{0}(\theta _{0}^{a})|d\theta _{0}^{a}|.
\label{d1}
\end{equation}
The summation in Eq.(\ref{d1}) is performed over all branches of the
function $\theta _{0}(\theta )$ defined by (\ref{mapping}). The modulus sign
is needed to cover the branches in which $\theta $ is a decreasing function
of $\theta _{0}$. The time-dependent angular distribution of the ensemble of
rotors is then given by 
\begin{equation}
f(\theta ,\tau )=\sum_{a}\frac{f(\theta _{0}^{a},\tau =0)}{\left| d\theta
/d\theta _{0}\right| _{a}}  \label{cldist}
\end{equation}
It follows immediately from Eq.(\ref{cldist}) that even for a smooth initial
distribution, the function $f(\theta ,\tau )$ exhibits a singular behavior
near the angles where $d\theta /d\theta _{0}=0.$ This condition (together
with Eq.(\ref{mapping})) defines the critical initial angle $\cos \overline{%
\theta _{0}} = 1/(P \tau)$ for which two classical trajectories coalesce at the same
final angle $\theta _{r}$ (see Fig. \ref{clmap3} (a)). We may call such an
event a rainbow-like catastrophe in analogy with the similar optical
phenomenon. As is clear from Fig. 2, two such rainbows appear at $%
\tau =\tau _{f}$, and they persist forever after that. Fig.~2
shows focus and rainbow singularities for the uniform initial angular
distribution. The time-dependent rainbow angle $\theta _{r}$ is given by 
\[
\theta _{r}=-\arccos (1/P\tau )+\sqrt{(P\tau )^{2}-1} 
\]
The other rainbow can be found at $-\theta _{r}$.

In the case of polarization-type interaction Eq.(\ref{inddipole}) we simply
replace Eq.(\ref{mapping}) by 
\[
\theta =\theta _{0}-P\tau \sin (2\theta _{0})\text{ \ \ \ \ \ \ \ \ \ \ \ \
\ \ }(%
\mathop{\rm mod}%
2\pi ) 
\]
Similarly to the previous case, the focusing happens at $\tau _{f}=1/(2P).$
However, it can be easily seen that the focusing occurs simultaneously at $%
\theta =0$ and $\theta =\pi $ now. \ Because of the symmetry of the angular
potential, two pairs of rainbows appear in the angular distribution at $\tau
>\tau _{f}$ .

\subsection{Classical description of an ensemble of 3D kicked rotors.}

We use spherical coordinates in which field direction is taken as $z$-axis,
and the angles $\theta ,\varphi $ are limited by $0\leqslant \theta
\leqslant \pi ,$ $\ 0\leqslant \varphi \leqslant 2\pi $. We restrict
ourselves by axially symmetric distributions only. For the dipole-type
coupling, the trajectories of the kicked rotors are obtained from Eq.(\ref
{mapping}) using the above restrictions for $\theta $ and $\theta _{0}$ (see
Fig.\ref{clmap3} (b)). The time-dependent distribution function obeys the
rule similar to (\ref{cldist}): 
\begin{equation}
f(\theta ,\tau )=\sum_{a}\frac{f(\theta _{0}^{a},\tau =0)\sin (\theta
_{0}^{a})}{\left| d\theta /d\theta _{0}\right| _{a}\sin (\theta )}
\label{3Ddist}
\end{equation}
Here the sum runs again over the branches of the function $\theta
_{0}(\theta )$ shown at Fig.\ref{clmap3}. Clearly, we obtain focusing at $%
\tau =\tau _{f}$ and a rainbow-like singularity at $\tau >\tau _{f}$
(whenever $\left| d\theta /d\theta _{0}\right| _{a}=0$). The rainbow first
appears at the north pole ($\theta =0$) and takes a shape of a ring moving
on the sphere toward the south pole. The rainbow ring shrinks to a point at
time $\tau _{f}^{\prime }$ defined from 
\[
\pi =-\arccos (\frac{1}{P\tau _{f}^{\prime }})+\sqrt{(P\tau _{f}^{\prime
})^{2}-1} 
\]
After that, the rainbow changes the direction of motion and runs toward the
north pole. These cycles repeat over and over again.

Comparing Eq.(\ref{cldist}) and Eq.(\ref{3Ddist}), we may identify another
source of singularities in the distribution function in 3D. Indeed, $%
f(\theta ,\tau )$ becomes singular at $\theta =0$ or $\theta =\pi $ whenever 
$\sin (\theta )=0$, provided the corresponding $\theta _{0}^{a}$ (called $%
\theta _{g}$) has a non-zero sinus-value. We call this singularity {\em glory%
} in analogy with the phenomenon well known in atmospheric optics and
quantum-mechanical scattering theory \cite{Ford,Berry}. As it will be shown
latter, the classical singularities are replaced by pronounced peaks
accompanied by fast oscillations in the quantum regime.

As follows from Eq.(\ref{mapping}), the glory appears at $\theta =0$ when
the equation 
\[
\theta _{g}=P\tau \sin (\theta _{g}) 
\]
has a non-trivial solution (see Fig.\ref{clmap3} (b)). The glory appears
first exactly at the focusing time $\tau _{f}=1/P$, and stays forever after
that. The glory peak of the distribution at the north pole is shown at Fig.%
\ref{orient}. Note that both $\overline{\theta _{0}}$ and $\theta _{g}$ are
zero at $P\tau =1.$ The {\em backward glory} appears at $\theta =\pi $ after
the rainbow ring reaches the south pole at $\tau =\tau _{f}^{\prime }$. For $%
\tau >\tau _{f}^{\prime }$ , two glory angles $\theta _{g1}^{\prime },\theta
_{g2}^{\prime }$ exist for which 
\[
\pi =\theta _{gs}^{\prime }-P\tau \sin (\theta _{gs}^{\prime })\text{ \ \ \
\ \ \ }s=1,2 
\]
After being created, the backward glory stays forever in the classical
model. Figure \ref{align} shows similar phenomena for the polarization-type
interaction.

\bigskip

\subsection{\protect\bigskip Quantum kicked \ rotor (2D)}

We proceed now to the quantum description of the kicked rotor, starting with
the simplest two-dimensional case. The dynamics of a quantum 2D rotor after
its interaction with a strong laser pulse was described in detail in \cite
{rotor,Lei02}. We reproduce briefly the relevant results here.

The wave function of a 2D rotor can be expanded in terms of eigenfunctions
of a free rotator 
\begin{equation}
\Psi (\theta ,\tau )=\frac{1}{\sqrt{2\pi }}\sum\limits_{m=-\infty }^{+\infty
}c_{n}(\tau )\exp (in\theta )  \label{f2}
\end{equation}
(here we use the dimensionless units). In the absence of the field, the wave
function is 
\begin{equation}
\Psi (\theta ,\tau )=\frac{1}{\sqrt{2\pi }}\sum\limits_{m=-\infty }^{+\infty
}c_{n}(0)\exp (-i\frac{n^{2}}{2}\tau +in\theta )  \label{f2free}
\end{equation}
Even in this simplest case, the wave function (\ref{f2free}) shows an
extremely rich space-time dynamics. In particular, $\Psi (\theta ,\tau )$
demonstrates a periodic behavior in time (quantum revivals) with the
''quantum'' period $T_{rev}=4\pi $\ \ (full revival), and a number of
fractional revivals at \ $\tau =p/s$\ $T_{rev}$ ($p$ and $s$ are mutually
prime numbers) \cite{Ave89a}.

The result of an extremely short pulse action on the quantum system can be
obtained by neglecting the effect of the kinetic energy term in the
Hamiltonian (\ref{3drotor}) during the pulse. The reduced Schr\"{o}dinger
equation can be immediately solved, providing the following rule connecting
the wave function before and after the kick: 
\begin{equation}
\Psi (\theta ,\tau _{k}+0)=\exp [-\frac{i}{\hbar}
\int d\tau \
V(\theta ,\tau )]\ \Psi (\theta ,\tau _{k}-0)  \label{connection}
\end{equation}
Here $\tau _{k}$ is the moment when the kick is applied, and integration is
performed over the short kick duration. In the case of the dipole-
type interaction, the exponent in Eq.(\ref{connection}) takes the form of $%
\exp [iP\cos (\theta )].$ Using Eq.(\ref{f2}), we derive the transformation
rule for the coefficients $c_{n}$: 
\[
c_{n}(\tau _{k}+0)=\sum\limits_{m=-\infty }^{+\infty
}i^{n-m}J_{n-m}(P)c_{n}(\tau _{k}-0) 
\]
Here $J_{n}(P)$ is Bessel function of the $n$th order. If, for example, we
apply a kick to a rotor being in its ground state at $\tau =0$ (i.e. $\Psi
(\theta ,0)=1/\sqrt{2\pi }$), then the following rotational wave packet is
created: 
\begin{equation}
\Psi (\theta ,\tau )=\sum\limits_{m=-\infty }^{+\infty }\frac{1}{2\pi }%
i^{n}J_{n}(P)\exp (-i\frac{n^{2}}{2}\tau +in\theta )  \label{onekick}
\end{equation}
Figure \ref{2Dq} shows time evolution of the probability density $|\Psi
(\theta ,\tau )|^{2}$ numerically calculated according to Eq.(\ref{onekick})
for a relatively large kick strength ($P=85$). We may identify there some
distinct phenomena that were already discussed in the classical context
above. First of all, the wave function shows an extreme narrowing in the
region $\theta \approx 0$ (Fig.\ref{2Dq} (b) ). Clearly, this effect may be
interpreted as a focusing at $\tau _{f}=1/P$. Immediately after the
focusing, two sharp singularities are formed in the distribution, which are
moving with time. They may be identified with the rainbow-type catastrophe
described above. The quantum nature of rotational motion replaces the
classical singularities by sharp maxima with the Airy-like shape typical to
the rainbow phenomenon (Fig.\ref{2Dq} (c)). We will discuss this in more
detail in the following sections presenting the semiclassical description of
the rotational wave packet dynamics.

It should be stressed, however, that the long-time regime of the quantum
motion is radically different from the classical one. Thus, contrary to the
classical limit in which the singular caustics exist forever, they gradually
disappear in the quantum case because of the overall dephasing of the
initial rotational wave packet. Figures \ref{2Dq} (d) - \ref{2Dq} (i) show
several examples of fractional foci and rainbows in the angular
distribution, which is a purely quantum effect.

The results for a 2D rotor with polarization-type interaction are very
similar to the ones given above, except the obvious symmetry differences. 
Therefore, we
proceed to the analysis of the three-dimensional effects now.

\subsection{\protect\bigskip Quantum kicked rotor (3D).}

The treatment of the three-dimensional case is rather similar to the previous
one, and we will briefly summarize the main differences. First of all, for a
vertically polarized (along $z$-axis) external field, the spherical
harmonics $Y_{l}^{0}(\Omega )$ with $\Omega=(\varphi,\theta)$
replace the eigenfunctions $\exp (in\theta )$%
\ in the rotational wave packet superposition. Next, the energy spectrum is $%
l(l+1)/2$ instead of $n^{2}/2$ now. A wave function of a molecule with
permanent dipole moment undergoes the following transformation as the result
of a $\delta $-pulse applied at $\tau =0$:

\begin{equation}
\psi (\Omega ,0^{+})=\exp \left( iP\cos \theta \right) \psi (\Omega ,0^{-})= 
\left[ \sum_{l=0}^{\infty }i^{l}\sqrt{4\pi (2l+1)}j_{l}(P)Y_{l}^{0}(\Omega )%
\right] \ \psi (\Omega ,0^{-})  \label{expcos}
\end{equation}
Here, $\ j_{l}(P)$ is spherical Bessel function of the order $l$, and $\psi
(\Omega ,0^{-}),\psi (\Omega ,0^{+})$ are the wave functions just before and
just after the interaction, respectively. If the rotor is initially in the
ground state, $\psi (\Omega ,0^{-})=Y_{0}^{0}=1/\sqrt{4\pi }$, then the wave
function at time $\tau $ after the kick becomes 
\begin{equation}
\psi (\theta ,\tau )=\sum_{l=0}^{\infty }i^{l}\sqrt{2l+1}\,j_{l}(P)\exp %
\left[ \frac{i}{2}l(l+1)\tau \right] Y_{l}^{0}(\theta ).  \label{dipoleqm}
\end{equation}
Fig. \ref{orient} (a), bottom line, shows the probability distribution $%
|\psi (\theta ,\tau )|^{2}$ for a strongly kicked ($P=75$) rotor at the
focusing time $\tau =1/P$ (a), and also at $\tau =2/P$ (b), and $\tau =4/P$
(c). The presence of a glory at $\theta =0$ and of rainbow with Airy-like
oscillations is clearly seen in these pictures. An analytical semiclassical
description of these features will be given in the following sections of the
paper.

If the external field couples with induced polarization of the molecule (see
Eq.(\ref{inddipole})), the wave function immediately after the kick can be
written as

\begin{equation}
\psi (\Omega ,0^{+})=\exp \left( iP\cos ^{2}\theta \right) \psi (\Omega
,0^{-})=\left[ \sqrt{4\pi }\exp \left( i\frac{P}{2}\right)
\sum_{l}c_{l}Y_{2l}^{0}(\theta )\right] \ \psi (\Omega ,0^{-}),
\end{equation}
with 
\begin{equation}
c_{l}=\frac{1}{\sqrt{4l+1}}\sum_{L}i^{L}j_{L}\left( \frac{P}{2}\right)
d_{L,l}.
\end{equation}
The coefficients $d_{L,l}$ can be determined by a recurrence relation (see
Appendix A). For a rotor initially in the ground state, the wave function
after a kick can be expressed as 
\begin{equation}
\psi (\theta ,\tau )=\frac{1}{\sqrt{4\pi }}\sum_{l}c_{l}\exp \left[
-il(2l+1)\tau \right] Y_{2l}^{0}(\theta ).
\label{cos2wf}
\end{equation}
(Note that only even spherical harmonics contribute to the wave function
here). Figure \ref{align} shows the probability distribution $|\psi (\theta
,\tau )|^{2}$ according to Eq.(\ref{cos2wf}) for $\tau =0.5/P$ (a), $\tau
=2/P$ (b), and $\tau =2.4/P$ (c), compared to the classical distribution
function. As in the case of molecules with a permanent dipole moment, we
observe focusing (a), rainbow and glory formation ((b) and (c)).

\bigskip

\section{Semiclassical approximations}

We have observed a close correspondence between the classical angular
distribution function and exact quantum mechanical probability distribution
in the case of strong kicks. This invites to a quasiclassical treatment of
the time-dynamics of a strongly kicked rotor. The following part of the
paper presents an asymptotic quasiclassical theory of the above critical
phenomena, which is based on integral representation of the rotor wave
function with the help of quantum mechanical propagator. For simplicity, the
discussion will be focused on rotors (molecules) having a permanent dipole
moment.

\subsection{Propagator for a 2D rigid rotor}

The propagator for a free quantum mechanical 2D rigid rotor is 
\begin{equation}
U(\theta _{0},0;\theta ,\tau )=\frac{1}{\sqrt{2i\pi \tau }}\sum_{p=-\infty
}^{\infty }\exp \left[ i\frac{(\theta -\theta _{0}+2\pi p)^{2}}{2\tau }%
\right] .
\end{equation}
The sum over $p$ corresponds to an infinite number of trajectories going
from $\theta _{0}$ to $\theta $ by making various number of full rotations.
The wave function of the rotor is therefore 
\begin{equation}
\psi (\theta ,\tau )=\int_{0}^{2\pi }d\theta _{0}U(\theta _{0},0;\theta
,\tau )\psi (\theta _{0},0).
\end{equation}
Since $\psi (\theta _{0}+2p\pi ,0)=\psi (\theta _{0},0)$, the wave
function can be written as 
\begin{equation}
\psi (\theta ,\tau )=\frac{1}{\sqrt{2i\pi \tau }}\int_{-\infty }^{\infty
}d\theta _{0}\exp \left[ i\frac{(\theta -\theta _{0})^{2}}{2\tau }\right]
\psi (\theta _{0},0).
\end{equation}
If interaction with the laser field is proportional to $\cos \theta $ (see
Eq.(\ref{permanent})), the wave function after the kick is 
\begin{equation}
\psi (\theta ,\tau )=\frac{1}{2\pi \sqrt{i\tau }}\int_{-\infty }^{\infty
}d\theta _{0}\exp \left[ i\Phi (\theta ,\theta _{0})\right]  \label{wf2d}
\end{equation}
with 
\begin{equation}
\Phi (\theta ,\theta _{0})=P\cos \theta _{0}+\frac{(\theta -\theta _{0})^{2}%
}{2\tau }.  \label{wf2d_phase}
\end{equation}

\subsection{A planar model for a 3D rigid rotor}

Classical considerations about the focusing process and the formation of
glory and rainbow suggest that those phenomena arise from particles that
have been initially in the region of small angles ($\theta <<1$). Therefore,
we expect to get a certain insight by considering an almost planar motion of
a rotor in the polar region. For this, we consider a flat 2D model in which
the wave function is initially uniformly distributed inside a circle of
area $4\pi$ and is kicked radially towards the center of the circle by a
cosine kick. 

According to this model, the wave function at time $\tau $ after the kick
can be written as 
\[
\psi (\theta ,\tau )=\frac{1}{4i\tau \pi ^{3/2}}\exp \left[ i\left( \frac{%
\theta ^{2}}{2\tau }+P\right) \right] I, 
\]
where $I$ is defined as 
\begin{equation}
I=\int_{0}^{2}d\theta _{0}\theta _{0}\int_{0}^{2\pi }d\varphi _{0}\exp 
\left[ i\Phi (\theta ,\theta _{0},\varphi _{0})\right] ,  \label{semiclprop}
\end{equation}
with the phase 
\begin{equation}
\Phi (\theta ,\theta _{0},\varphi _{0})=P\frac{\theta _{0}^{4}}{24}+\frac{1}{%
2}\left( \frac{1}{\tau }-P\right) \theta _{0}^{2}-\frac{\theta \theta _{0}}{%
\tau }\cos \varphi _{0}.  \label{planarphase}
\end{equation}
Since we assume that only small values of $\theta _{0}$ are important, we
expanded $\cos \theta _{0}$ up to fourth order. Integrating Eq.({\ref
{semiclprop}) over $\varphi _{0}$, we find 
\begin{equation}
\psi (\theta ,\tau )=\frac{1}{i\tau \sqrt{4\pi }}\exp \left[ i\left( P+\frac{%
\theta ^{2}}{2\tau }\right) \right] \int_{0}^{2}d\theta _{0}\theta
_{0}J_{0}\left( \frac{\theta \theta _{0}}{\tau }\right) \exp \left[ i\left( 
\frac{\theta _{0}^{2}}{2}\left( \frac{1}{\tau }-P\right) +\frac{P}{24}\theta
_{0}^{4}\right) \right] ,  \label{semicl_2}
\end{equation}
where $J_{0}$ is Bessel function of zero-th order. Figure \ref{pmod} shows
the probability distribution $|\psi (\theta ,\tau )|^{2}$ for a fully
three-dimensional rotor system (Eq.(\ref{dipoleqm})) compared to the
probability distribution according to Eq.(\ref{semicl_2}) at different
times. It can be seen, that for small $\tau $, the planar model describes
the dynamics of the rigid rotor quite accurately. Focusing (Fig. \ref{pmod}%
(a)) and glory formation (Figs. \ref{pmod}(b) and (c)) appear also in the
planar model. Only at larger times (see Fig. \ref{pmod}(d)), when the wave
function of the 3D rotor depends on contributions from larger angles $\theta 
$, the planar model fails. Moreover, by analytically calculating the wave
function at the focal point (see Appendix B), we show that the wave function
in the planar model has the right asymptotic behavior for large $P$.
In Sec. IV.B, we will use the planar model exactly as it was defined
above to describe the focusing of the wave packet which occurs at zero
angle and for short time. In Sections V.B and VI, an extension of the model
will be practiced since we will use the full $P \cos \theta_0$ term inside the
phase. Although not justified for the large angles that we will consider, we
feel that the physics of the planar model is sound enough to provide a
qualitative understanding of the behavior of the wave packet on the
sphere for not too large times.


\subsection{Topology of stationary points}

For a large kick strength $P\gg 1$, the exponent in Eq.~(\ref{wf2d}) and
Eq.~(\ref{semiclprop}) is a fast-oscillating function. Therefore, the main
contribution to the integrals comes from the points of stationary phase,
where $\partial \Phi /\partial \theta _{0}=0$ (and $\partial \Phi /\partial
\varphi _{0}=0$ in the 3D case). Figure \ref{clmap3} shows the points of
stationary phase for every angle $\theta $ for (a) 2D and (b) 3D rotors. The
shape of the wave function is determined by the topology of the points of
stationary phase. The simplest semiclassical approximation is the stationary
phase approximation, which can be used if the stationary points are well
separated. However, the semiclassical catastrophes like rainbow and glory
formation occur when the stationary points are close to each other and,
even, merge.

Thus, if $P\tau \approx 1$, the three points of stationary phase around $%
\theta \approx 0$ (and for $\theta \approx 2\pi $ in the 2D case) are
non-isolated (cusp). For $P\tau \rightarrow 1$, they merge into a single
one, giving rise to focusing. In this region, all three stationary points
have to be treated together. In Sec.~IV we show how the wave function can be
approximated with the help of Pearcey approximation in this case.

Next, we consider the case of $P\tau \geq 1$. For $\theta \approx
\theta _{r}$, we have two close stationary points merging into a single one
at $\theta =\theta _{r}$. This fold-type catastrophe causes the formation of
angular rainbow which can be treated with the help of Airy approximation, as
we will show in Sec. V. In three-dimensional case, the points of stationary
phase belong to two different branches (see Fig.\ref{clmap3} (b)). The point
with $\theta _{0}>\theta _{g}$ corresponds to $\varphi _{0}=0$, while the
two points with $\theta _{0}<\theta _{g}$ belong to the branch with $\varphi
_{0}=\pi $. These two branches merge at $\theta _{0}=\theta _{g}$ giving
rise to the glory formation, which is absent in the 2D case. In Sec. VI we
show how this phenomenon can be treated with the help of Bessel
approximation.

\section{Focusing}

\subsection{Focusing of a 2D rotor}

In this section we consider 2D kicked rotor close to focusing ( $P\tau
\approx 1$). As can be seen in Fig.~\ref{clmap3} (a), there are three close
points of stationary phase near $\theta \approx 0$, and three stationary
points at $\theta \approx 2\pi $ in this case. Due to the symmetry, it is
sufficient to consider only the stationary points at $\theta \approx 0$. The
complete wave function Eq.(\ref{wf2d}) can be written as 
\begin{equation}
\psi (\theta ,\tau )=\tilde{\psi}(\theta ,\tau )+\tilde{\psi}(2\pi -\theta
,\tau ).
\end{equation}
In order to account for the three stationary points, we keep all terms up to
the order $\theta _{0}^{4}$ in the phase Eq.(\ref{wf2d_phase}). Then, 
\begin{equation}
\tilde{\psi}(\theta ,\tau )\approx \frac{1}{\sqrt{2i\pi \tau }}\exp \left[
i\left( \frac{\theta ^{2}}{2\tau }+P\right) \right] \,\frac{1}{\sqrt{2\pi }}%
\int_{-\infty }^{\infty }\exp \left\{ i\left[ \frac{\theta _{0}^{4}}{24}P+%
\frac{\theta _{0}^{2}}{2}\left( \frac{1}{\tau }-P\right) -\frac{\theta
\theta _{0}}{\tau }\right] \right\} \,d\theta _{0}\;.  \label{2dfwp}
\end{equation}
Making the following substitutions 
\[
u=\left( \frac{P}{24}\right) ^{1/4}\theta _{0},\quad \quad x=\left[ \frac{6}{%
P}\right] ^{\frac{1}{2}}\left( \frac{1}{\tau }-P\right) ,\quad \mbox{and}%
\quad \beta =\sqrt{2}\,\frac{\theta }{\tau }\left[ \frac{6}{P}\right] ^{%
\frac{1}{4}}\;, 
\]
we can write the wave function Eq.(\ref{2dfwp}) as 
\begin{equation}
\tilde{\psi}(\theta ,\tau )=\frac{1}{\pi \sqrt{2i\tau }}\left[ \frac{6}{P}%
\right] ^{\frac{1}{4}}\exp \left[ i\left( \frac{\theta ^{2}}{2\tau }%
+P\right) \right] \,{\cal P}(x,\beta )\;,  \label{2dfwp1}
\end{equation}
where ${\cal P}$ is the Pearcey function 
\begin{eqnarray}
{\cal P}(x,\beta ) &=&{\cal P}(x,-\beta )~=~\int_{-\infty }^{\infty }\exp
[i(u^{4}+xu^{2}+\beta u)]\,du  \label{pearcey} \\
&=&\frac{1}{2}\sum_{n=0}^{\infty }\sum_{m=0}^{\infty }\frac{x^{m}}{m!}\frac{%
\beta ^{2n}}{2n!}\Gamma \left[ \frac{1}{4}(2n+2m+1)\right] \exp \left[ i%
\frac{\pi }{8}(10n+6m+1)\right] \;.  \nonumber
\end{eqnarray}
This function has been studied intensively (see, e.g. \cite{con73,Ber80a}.
If $P\tau =1$, that is, at focal time, then $x=0$, and the wave function Eq.(%
\ref{2dfwp1}) can be reduced to a single sum, 
\begin{eqnarray}
\tilde{\psi}\left( \theta ,\frac{1}{P}\right) &=&\frac{1}{2\pi \sqrt{2i}}%
(6P)^{1/4} \exp \left[ iP\left( 1+\frac{\theta ^{2}}{2}\right) \right] 
\nonumber \\
&\times &\sum_{n=0}^{\infty }\frac{\beta ^{2n}}{2n!}\Gamma \left[ \frac{1}{4}%
(2n+1)\right] \exp \left[ i\frac{\pi }{8}(10n+1)\right] .  \label{2dwfft}
\end{eqnarray}
At the origin $\theta =0$ the probability distribution is 
\begin{equation}
\left| \tilde{\psi}\left( 0,\frac{1}{P}\right) \right| ^{2}=\frac{1}{8\pi
^{2}}\sqrt{6P} \,\Gamma ^{2}\left( \frac{1}{4}\right) \approx 0.4078\sqrt{P}.
\end{equation}
For $P\rightarrow \infty $, the probability distribution at $\theta =0$ goes
to infinity, which is consistent with the classical picture. Moreover, we
can determine from Eq.(\ref{2dwfft}) that for large angles $\theta $, the
probability distribution is independent of $P$ 
\begin{equation}
\left| \tilde{\psi}\left( \theta ,\frac{1}{P}\right) \right| ^{2}\rightarrow 
\frac{1}{\pi (6\theta )^{2/3}}.
\end{equation}
Considering the classical angular distribution function of a 2D rotor (see 
\cite{rotor,Lei02}), 
\begin{equation}
f(\theta ,1/P)=\frac{1}{2\pi }\frac{1}{|d\theta /d\theta _{0}|},
\end{equation}
with $\theta (\tau )=\theta _{0}-P\tau \sin (\theta _{0})\,(\mbox{mod}\,2\pi
)$, we obtain the same result by expanding $\sin \theta _{0}$ up to the
cubic term.

In Fig. \ref{2dim7} the wave function in the Pearcey approximation Eq.(\ref
{pearcey}) is compared to the exact wave function. Fig. \ref{2dim7} (a)
demonstrates that the focusing is very accurately described in the Pearcey
approximation. For $\tau =1.1/P$ (b) and $\tau =1.2/P$ (c), we still observe
a good agreement with the exact quantum mechanical result, but for larger
times (see $\tau =1.4/P$ in Fig. \ref{2dim7} (d) ), the approximation starts
to break down.

\subsection{Focusing of a 3D rotor}

We proceed now to the case of $P\tau \approx 1$ for a 3D rotor (see Fig.\ref
{clmap3} (b)). In order to treat all three points of stationary phase around 
$\theta =0$ together, we have to keep the full quartic expression in the
phase $\Phi (\theta ,\theta _{0},\varphi _{0})$ in Eq.(\ref{planarphase}) to
accurately calculate the integral. Due to the geometrical factor $\theta
_{0} $ in front of the exponent in integral Eq.(\ref{semiclprop}), the
integration over $\theta _{0}$ can be expressed with help of a derivative of
Pearcey integral. We can write the wave function as 
\begin{equation}
\psi (\theta ,\tau )=-\frac{1}{4\tau \pi ^{3/2}}\exp \left[ i\left( P+\frac{%
\theta ^{2}}{2\tau }\right) \right] \left( \frac{24}{P}\right)
^{1/2}\int_{0}^{2\pi }d\varphi _{0}\frac{\partial {\cal P}_{1}(x,y)}{%
\partial y},  \label{airy}
\end{equation}
where 
\begin{equation}
{\cal P}_{1}(x,y)=\int_{0}^{\infty }du\exp \left[ i\left(
u^{4}+xu^{2}+yu\right) \right]
\end{equation}
is a part of Pearcey integral, such that ${\cal P}(x,y)={\cal P}_{1}(x,y)+%
{\cal P}_{1}(x,-y)$. Here, $y=-\beta \cos \varphi _{0}$. According to \cite
{con81}, the integral ${\cal P}_{1}$ can be represented as a sum: 
\begin{equation}
{\cal P}_{1}=\frac{1}{4}\sum_{n=0}^{\infty }\sum_{m=0}^{\infty }\frac{x^{m}}{%
m!}\frac{y^{n}}{n!}\Gamma \left[ \frac{1}{4}(n+2m+1)\right] \exp \left[ 
\frac{1}{8}i\pi \left( n+6m+1\right) \right] \exp \left( i\frac{1}{2}\pi
n\right) .
\end{equation}
Since 
\begin{equation}
\int_{0}^{2\pi }d\varphi _{0}\cos ^{n-1}\varphi _{0}=2\pi \frac{(n-2)!!}{%
(n-1)!!}
\end{equation}
for odd $n$ and zero if $n$ is even, the wave function Eq.(\ref{airy}) can
be written as 
\begin{eqnarray}
\psi (\theta ,\tau ) &=&-\left( \frac{6}{P}\right) ^{1/2}\exp \left[ i\left(
P+\frac{\theta ^{2}}{2\tau }\right) \right] \frac{1}{4\sqrt{\pi }\tau }%
\sum_{n=0}^{\infty }\sum_{m=0}^{\infty }\frac{x^{m}}{m!}\frac{\beta ^{2n}}{%
(2n)!}\frac{(2n-1)!!}{(2n)!!}  \nonumber \\
&\times &\Gamma \left[ \frac{1}{2}(n+m+1)\right] \exp \left[ i\frac{\pi }{4}%
(5n+3m+3)\right] .  \label{cusp1}
\end{eqnarray}
Comparison of the results obtained with this expression with those resulting
from the exact quantum evolution for $P\tau \in \lbrack 1,1.4]$ and $\theta
\in \lbrack 0,0.3]$ is presented in Fig. \ref{3dim7}. As in the 2D case, we
can see that Pearcey approximation describes the focusing and glory very
well at times $P\tau \approx 1$. All three stationary points merge into a
single one $\theta _{0}=0$ at focusing time $P\tau =1$. In this case, $x=0$,
and Eq.({\ref{cusp1}) can be simplified to 
\begin{equation}
\psi \left( \theta ,\frac{1}{P}\right) =-(6P)^{1/2}\exp \left[ i\left( P+%
\frac{\theta ^{2}}{2\tau }\right) \right] \frac{1}{4\sqrt{\pi }}%
\sum_{n=0}^{\infty }\frac{\beta ^{2n}}{(2n)!}\frac{(2n-1)!!}{(2n)!!}\Gamma %
\left[ \frac{1}{2}(n+1)\right] \exp \left[ i\frac{\pi }{4}(5n+3)\right] .
\label{cusp2}
\end{equation}
Fig. \ref{3dim7} (a) shows the focused wave function according to Eq.(\ref
{cusp2}) compared to the exact quantum result. At the origin $\theta =0$,
wave function Eq.(\ref{cusp2}) is 
\begin{equation}
\left| \psi \left( 0,\frac{1}{P}\right) \right| ^{2}=\frac{3 P}{8 \pi} 
 \,\Gamma
^{2}\left( \frac{1}{2}\right) =\frac{3}{8}P.
\end{equation}
As in the 2D case, the probability distribution diverges as $P\rightarrow
\infty $. However, in the 3D case, the probability distribution is
proportional to $P$ , while in the 2D case, it is proportional to $\sqrt{P}$%
. }

Comparing Fig.\ref{2dim7} (c) and (d) with Fig.\ref{3dim7} (c) and (d), one
can recognize the difference between the 2D and 3D cases. In the former case
the peak at $\theta =0$ splits into a symmetric pair of peaks moving away
from $\theta =0$ with time. In the 3D case, the peak stays at $\theta =0$
after being created. This is the result of the glory effect mentioned before
in the classical context, and discussed in detail in one of the following
sections.

\section{Rainbow}

\subsection{2D rotational wave packet in terms of Airy functions}

In the following, we investigate the formation of angular rainbow, that is,
the shape of wavefunction around the rainbow angle for $P\tau >1$. In the 2D
case, we can approximate the wave packet Eq.(\ref{wf2d}) as a sum of two
contributions related to the rainbow angles $\theta =\theta _{r}$ and $%
\theta =2\pi -\theta _{r}$ (see Fig.~\ref{clmap3} (a)). Due to the symmetry
we can write 
\begin{equation}
\psi (\theta ,\tau )\approx \psi _{\theta _{r}}(\theta ,\tau )+\psi _{\theta
_{r}}(2\pi -\theta ,\tau )\;.  \label{2wpapp}
\end{equation}
The rainbow angle $\theta _{r}$ is defined as $\theta _{r}=\bar{\theta _{0}}%
-P\tau \sin \bar{\theta _{0}}$, and the phase Eq.(\ref{wf2d_phase}) can be
expanded around $\theta _{r}$ as 
\begin{equation}
\Phi (\theta ,\theta _{0})=\frac{2+(\theta -\bar{\theta}_{0})^{2}}{2\tau }+%
\frac{\theta _{r}-\theta }{\tau }(\theta _{0}-\bar{\theta}_{0})+P\sin \bar{%
\theta _{0}}\frac{(\theta _{0}-\bar{\theta _{0}})^{3}}{3!}+\ldots .
\label{expansion}
\end{equation}
Using the definitions 
\begin{equation}
y\equiv \left[ \frac{P\sin \bar{\theta _{0}}}{2}\right] ^{1/3}(\theta _{0}-%
\bar{\theta _{0}})\quad \mbox{and}\quad \eta \equiv \left[ \frac{2}{P\sin 
\bar{\theta _{0}}}\right] ^{1/3}\frac{(\theta _{r}-\theta )}{\tau }
\label{def}
\end{equation}
the phase $\Phi $ can be approximated by 
\[
\Phi (\theta ,\theta _{0})\simeq \frac{2+(\theta -\bar{\theta _{0}})^{2}}{%
2\tau }+\eta y+\frac{y^{3}}{3}. 
\]
Within this approximation 
\[
\psi _{\theta _{r}}(\theta ,\tau )\approx \frac{1}{2\pi \sqrt{i\tau }}\,\exp %
\left[ i\left( \frac{2+(\theta -\bar{\theta _{0}})^{2}}{2\tau }\right) %
\right] \left[ \frac{2}{P\sin \bar{\theta _{0}}}\right] ^{1/3}\,\int_{-%
\infty }^{+\infty }e^{i(\eta y+y^{3}/3)}\,dy. 
\]
Using formula $\mbox{Ai}(\eta )=\frac{1}{\pi }\int_{0}^{\infty }\cos
(y^{3}/3+\eta y)\,dy$ one obtains 
\begin{equation}
\psi _{\theta _{r}}(\theta ,\tau )=\frac{1}{\sqrt{i\tau }}\left[ \frac{2}{%
P\sin \bar{\theta _{0}}}\right] ^{1/3}\,\exp \left[ i\left( \frac{2+(\theta -%
\bar{\theta _{0}})^{2}}{2\tau }\right) \right] \,\mbox{Ai}(\eta ).  \label{H}
\end{equation}

Figure \ref{ffairy} illustrates the quality of approximating the 2D wave
packet by a sum of two contributions expressed through Airy functions around
rainbow angles Eq.(\ref{2wpapp}). We have chosen $P\tau =4,4.7\mbox{~and~}6$
in (a), (b) and (c), respectively. The case (a) corresponds to the rainbow
angle $\theta _{r}=2.555$ and the two wave structures moving on the circle
in opposite directions are still well separated. The interference between
them is so small that it is not seen in the figure. For case (b), $\theta
_{r}=3.236$ and the two semiclassical wave functions interfere strongly,
reproducing quite well the structure of the exact quantum wave packet. In
case (c), $\theta _{r}=4.513$ \ and the interference region extends between $%
\pi /2$ and $3\pi /2$ (more precisely, between two rainbow angles). Again,
the semiclassical wave function reproduces very well the qualitative
features of the quantum wave packet.

\subsection{Uniform Airy approximation for rainbow in 3D case}

In order to calculate the 3D wave function for $P\tau >1$ and $\theta
\approx \theta _{r}$, we have to estimate the double integral Eq.(\ref
{semiclprop}). Using stationary phase approximation for the integration over 
$\varphi _{0}$, we obtain 
\begin{equation}
I\approx \sqrt{\frac{-i\pi \tau }{2\theta }}\,\int_{0}^{\pi }d\theta
_{0}\,\theta _{0}^{\frac{1}{2}}\,\exp \left[ i\left( \frac{(\theta +\theta
_{0})^{2}}{2\tau }+P\cos \theta _{0}\right) \right] \;.  \label{e11}
\end{equation}
Here we consider only the azimuthal stationary point $\varphi _{0}=\pi $
that leads to two stationary points $\theta _{03}$ and $\theta _{02}$ in Fig.%
\ref{clmap3} (b). The second stationary azimuthal point $\varphi _{0}=0$ is
related to the isolated stationary point $\theta _{01}$ (Fig.\ref{clmap3}
(b)) and its contribution to the rainbow wave function is negligible. To
treat this important 3D case, we rely on the uniform WKB approximation that
is supposed to be superior compared to the plain asymptotic result based on
expansion Eq.(\ref{expansion}). The wave function in the uniform Airy
approximation is given by 
\begin{equation}
\psi _{\theta _{r}}^{(UA)}(\theta ,\tau )=\frac{1}{4i\tau \pi ^{\frac{3}{2}}}%
\,I\;.  \label{e17}
\end{equation}
Since the phase Eq.(\ref{planarphase}) close to the stationary points $%
\theta _{02}$ and $\theta _{03}$ has almost cubic structure (see (\ref
{expansion})), one can approximate $I$ \ in the region $\theta <\theta _{r}$
as the integral 
\begin{eqnarray}
J &=&\sqrt{\frac{-i\pi \tau }{2\theta }}\int_{-\infty }^{\infty }dx\,\frac{%
g_{1}+g_{2}x}{2\pi }\exp \left[ i(A+\xi x+x^{3}/3)\right]  \nonumber
\label{e15} \\
&=&\sqrt{\frac{-i\pi \tau }{2\theta }}\,e^{iA}\,\left[ g_{1}\mbox{Ai}(\xi
)-ig_{2}\mbox{Ai'}(\xi )\right] \;,
\end{eqnarray}
We fix parameters $g_{1}$, $g_{2}$, $A$, and $\xi $ in such a manner that
the values of $I$ and $J$ coincide at the points where their phase is
stationary. All these quantities depend on $\theta $. The stationary points
of $I$ satisfy $\xi ^{2}+x^{2}=0$. As $\xi <0$, $x_{1}=-\sqrt{-\xi }$ and $%
x_{2}=\sqrt{-\xi }$. From identification of the phases we finally obtain 
\begin{equation}
A=\frac{\Phi (\theta ,\theta _{03})+\Phi (\theta ,\theta _{02})}{2}
\end{equation}
and 
\begin{equation}
\xi =-\left[ \frac{3}{4}\left( \Phi (\theta ,\theta _{03})-\Phi (\theta
,\theta _{02})\right) \right] ^{2/3},
\end{equation}
where 
\begin{equation}
\Phi (\theta ,\theta _{0s})=\frac{(\theta +\theta _{0s})^{2}}{2\tau }+P\cos
\theta _{0s},
\end{equation}
with $s=1,2$. Identification of the integrals leads to two linear equations
for $g_{1}$ and $g_{2}$. Solving them, one obtains 
\begin{equation}
g_{1}=\pi \left( \frac{2}{P}\right) ^{\frac{1}{2}}\,|\xi |^{\frac{1}{4}%
}\left\{ \left[ \frac{\theta _{02}}{|\cos \bar{\theta}_{0}-\cos \theta _{02}|%
}\right] ^{\frac{1}{2}}+\left[ \frac{\theta _{03}}{|\cos \bar{\theta}%
_{0}-\cos \theta _{03}|}\right] ^{\frac{1}{2}}\right\}  \label{e16}
\end{equation}
and 
\begin{equation}
g_{2}=\pi \left( \frac{2}{P}\right) ^{\frac{1}{2}}\,|\xi |^{-\frac{1}{4}%
}\left\{ \left[ \frac{\theta _{02}}{|\cos \bar{\theta}_{0}-\cos \theta _{02}|%
}\right] ^{\frac{1}{2}}-\left[ \frac{\theta _{03}}{|\cos \bar{\theta}%
_{0}-\cos \theta _{03}|}\right] ^{\frac{1}{2}}\right\}  \label{e16a}
\end{equation}
For $\theta =\theta _{r}$ the coefficient $g_{2}$ approaches zero. Hence,
for $\theta \geq \theta _{r}$ the uniform Airy approximation reduces to Airy
approximation 
\begin{equation}
\Psi _{\theta _{r}}(\theta ,\tau )=\frac{1}{4\pi }\,\sqrt{\frac{i}{2\tau
\theta }}\,G_{1}\,e^{iA}\,\mbox{Ai}(\eta )\;,  \label{e19}
\end{equation}
where 
\begin{equation}
G_{1}=\lim_{\theta \rightarrow \theta _{r}}g_{1}=2\pi \left[ \frac{2}{P\sin 
\bar{\theta}_{0}}\right] ^{\frac{1}{3}}\sqrt{\bar{\theta}_{0}}  \label{e18}
\end{equation}
and $\eta $ is defined in (\ref{def}). Calculations performed according to
these formulas are presented in figure \ref{f8}. The uniform Airy (UA)
approximation gives quite good description of the quantum wave function. The
figures do not show the wave function shape and value for very small $\theta 
$ where the UA approximation fails (singularity $1/\sqrt{\theta }$). The
quantum wave function is normalized ($2\pi \int_{0}^{\pi }|\Psi (\theta
,\tau )|^{2}\sin \theta d\theta =1$). For the wave function in UA
approximation this integral gives 0.838. The main difference between the
Airy approximation for the 2D rotor and the UA approximation for the 3D
rotor is the $1/\sqrt{\theta }-$factor in the 3D case, which leads to the
divergence of the 3D semiclassical wave function at very small values of $%
\theta $. This singularity, however, does not show up in the probability
density as the latter is given by $2\pi |\psi (\theta ,\tau )|^{2}\sin
\theta $.

\section{Glory}

In the 3D case, for $P\tau \geqslant 1$ and $\theta \approx 0$, we have an
isolated stationary point around $\theta _{0}=0$, and two close stationary
points ($\theta _{02}$ and $\theta _{01}$) around the glory angle $\theta
_{g}$ (see Fig.\ref{clmap3} (b)). The latter pair of stationary points is
responsible for the formation of glory, and it can be treated by the uniform
Bessel approximation \cite{ber69}. According to \cite{ber69}, we perform the
integration over $\theta _{0}$ in Eq.(\ref{semiclprop}) by using the
ordinary stationary phase approximation: 
\begin{equation}
I\approx 2\sqrt{2\pi }\int_{0}^{\pi }d\varphi _{0}\frac{\theta _{0}(\varphi
_{0})}{\left| \partial ^{2}\Phi /\partial \theta _{0}^{2}\right| ^{1/2}}\exp %
\left[ i\left( \Phi \lbrack \theta _{0}(\varphi _{0})]-\chi \right) \right]
, 
\label{intbessel1}
\end{equation}
Here $\chi =-\pi /4$ if $\partial ^{2}\Phi /\partial \theta _{0}^{2}>0$ and $%
\chi =\pi /4$ in the opposite case. Angle $\theta _{0}(\varphi _{0})$ stands
for the stationary value of $\theta _{0}$ for a given value of $\varphi _{0}$%
. In order to perform the remaining integration, we map the phase $\Phi %
\left[ \theta _{0}(\varphi _{0})\right] $ with the function 
\begin{equation}
\Phi (\theta ,\psi )=a(\theta )-b(\theta )\cos \psi .
\end{equation}
The mapping is one-to-one if $\varphi _{0}=0$ corresponds to $\psi =0$, and $%
\varphi _{0}=\pi \ $\ to $\psi =\pi $. Then 
\begin{equation}
a(\theta )=\frac{1}{2}\left[ \Phi (\theta _{02})+\Phi (\theta _{01})\right]
\quad \mbox{and}\quad b(\theta )=\frac{1}{2}\left[ \Phi (\theta _{02})-\Phi
(\theta _{01})\right] .
\end{equation}
Here, $\theta _{01}$ and $\theta _{02}$ are the stationary points
corresponding to $\varphi _{0}=0$ and $\varphi _{0}=\pi $, respectively. The
integral in Eq.(\ref{intbessel1}) can be written as 
\begin{equation}
I=2\sqrt{2\pi \tau }\exp \left[ i\left( a(\theta )-\chi \right) \right]
\int_{0}^{\pi }d\psi \frac{\theta _{0}\frac{d\varphi _{0}}{d\psi }}{\left|
(1-P\tau )+\frac{P\tau }{2}\theta _{0}^{2}\right| ^{1/2}}\exp \left[
-ib(\theta )\cos \psi \right] .
\end{equation}
By differentiating $\Phi $ with respect to $\psi $, we obtain (using $%
\partial \Phi /\partial \theta _{0}=0$), 
\begin{equation}
\frac{\partial \Phi }{\partial \varphi _{0}}\frac{d\varphi _{0}}{d\psi }=%
\frac{\theta \theta _{0}}{\tau }\sin \varphi _{0}\frac{d\varphi _{0}}{d\psi }%
=b(\theta )\sin \psi .  \label{fdir}
\end{equation}
By taking another derivative of Eq.(\ref{fdir}) at $\psi =0$ \ and $\psi
=\pi $ we arrive at 
\begin{equation}
b(\theta )=\frac{\theta \theta _{0}}{\tau }\left( \frac{d\varphi _{0}}{d\psi 
}\right) ^{2}.
\end{equation}
which determines $d\varphi _{0}/d\psi $ at the stationary points. Following
the derivation of Berry \cite{ber69}, we define 
\begin{equation}
\frac{\theta _{0}\frac{d\varphi _{0}}{d\psi }}{\left| (1-P\tau )+\frac{P\tau 
}{2}\theta _{0}^{2}\right| ^{1/2}}\equiv p_{+}(\theta )+p_{-}(\theta )\cos
\psi .
\end{equation}
According to the mapping conditions, $p_{\pm }$ can be determined as 
\begin{equation}
p_{\pm }(\theta )=\frac{1}{2}\sqrt{\frac{b(\theta )\tau }{\theta }}\left[
\left( \frac{\theta _{01}}{(1-P\tau )+\frac{P\tau }{2}\theta _{01}^{2}}%
\right) ^{1/2}\pm \left( \frac{\theta _{02}}{(1-P\tau )+\frac{P\tau }{2}%
\theta _{02}^{2}}\right) ^{1/2}\right] .
\end{equation}
The integral can then be written as 
\begin{eqnarray}
I &=&2\sqrt{2\pi \tau }\exp \left[ i\left( a(\theta )-\chi \right) \right]
\int_{0}^{\pi }d\phi \left[ p_{+}(\theta )+p_{-}(\theta )\cos \psi \right]
\exp \left[ ib(\theta )\cos \psi \right]  \nonumber \\
&=&2\pi \sqrt{2\pi \tau }\exp \left[ i\left( a(\theta )-\chi \right) \right]
\left\{ p_{+}(\theta )J_{0}\left[ b(\theta )\right] -ip_{-}(\theta )J_{1}%
\left[ b(\theta )\right] \right\} .
\end{eqnarray}
The wave function is 
\begin{equation}
\psi (\theta ,\tau )=\frac{1}{i(2\pi )^{3/2}\tau \sqrt{2}}\exp \left[
i\left( \frac{\theta ^{2}}{2\tau }+P\right) \right] I.
\end{equation}
Figure ~\ref{bessel_fig} shows the modulus squared of the wave function (in
the uniform Bessel approximation) for $P=75$ and $\tau =4/P$ in comparison
with the exact numerical integration of Eq.(\ref{semicl_2}). For small
angles $\theta $, the quantum mechanical wave function is well represented
by the uniform Bessel approximation, and the formation of the angular glory
at $\theta \approx 0$ is reproduced. However, the approximation breaks down
if $\theta $ approaches the rainbow angle $\theta _{r}$.

If $\theta \rightarrow 0$, the stationary points $\theta _{01}$ and $\theta
_{02}$ merge to a single stationary point $\theta _{g}$. For time only
slightly larger than the focusing time, \ $\theta _{g}$ is defined by 
\begin{equation}
\theta _{g}=\sqrt{6\left( \frac{P\tau -1}{P\tau }\right) }
\end{equation}
and the wave function becomes 
\begin{equation}
\psi (\theta ,\tau )=\frac{1}{i\sqrt{2\tau }}\exp [i(a(\theta )-\chi )]\frac{%
\theta _{g}J_{0}\left( \frac{\theta _{g}\theta }{\tau }\right) }{\sqrt{%
1-P\tau +\theta _{g}^{2}\frac{P\tau }{2}}}\exp \left[ i\left( P+\frac{\theta
^{2}}{2\tau }+\frac{1}{2}\left( \frac{1}{\tau }-P\right) \theta _{g}^{2}+%
\frac{P}{24}\theta _{g}^{4}\right) \right] .
\end{equation}
This expression is analogous to Ford-Wheeler formula in nuclear scattering 
\cite{Ford}. \ It can be also derived by directly calculating the integral
in Eq.(\ref{semicl_2}) with the help of simple stationary phase
approximation using the glory angle $\theta _{g}$ as a single stationary
point.

\section{Kicked 3D rotor at finite temperature (classical treatment)}

In the previous sections, we considered a quantum rotor being initially in
its ground state, or, in the classical context, an ensemble of rotors with
zero initial velocity. In the present section we investigate the dynamics of
an ensemble of kicked classical rotors with a finite initial temperature,
and compare the results with the zero-temperature case.

The time evolution of a 2D rotor under thermal conditions was studied in 
\cite{Lei02}. In the previous Sections we have already observed several
differences between 2D and 3D kicked rotors at zero initial temperature.
Another difference appears at finite temperature because of the conservation
of the angular momentum projection onto the field polarization direction. In
a thermal ensemble, molecules generally have a non-zero initial projection
of this kind. As a result, an effective repulsive centrifugal force prevents
the rotors from reaching the exact $\theta =0$ and $\theta =\pi $
orientations. Therefore, two {\em holes} in the angular distribution of a
driven 3D thermal ensemble should be always present at $\theta =0$ and $%
\theta =\pi $ (see a related discussion in \cite{Zon1}).

In the following, we describe the dynamics of a classical ensemble of 3D
rotors under thermal conditions. The equation of motion for the angle $%
\theta $ of a free 3D rotor is given by 
\begin{equation}
\ddot{\theta}(t)=\frac{p^2_{\varphi }}{I^{2}}\frac{\cos \theta }{\sin
^{3}\theta },  \label{theta}
\end{equation}
where the canonical momentum 
$p_{\varphi}$ is a constant of motion.
We introduce dimensionless variables by defining the thermal
momentum $p_{th} = I \omega_{th}$ with $\omega_{th} = \sqrt{k_B T/I}$,
where $T$ is the temperature and $k_B$ the Boltzman constant.
Equation (\ref{theta}) can then be written as
\begin{equation}
\ddot{\theta}(t')=p'^2_{\varphi} \, \frac{\cos \theta }{\sin
^{3}\theta },  \label{thetadl}
\end{equation}
where $p'_{\varphi} = p_{\varphi}/p_{th}$ is the dimensionless momentum
and $t' = \omega_{th} t$ is the dimensionless time.
The differential equation Eq.(\ref{thetadl}) can be integrated by
substituting $u=\cos \theta$, so that
\begin{equation}
\cos \theta(t') = \frac{1}{2} \left(1 - \frac{p'_\theta}{\omega} \right)
\cos \left [\theta(0) + \omega t' \right] + 
\frac{1}{2} \left(1 + \frac{p'_\theta}{\omega} \right)
\cos \left [\theta(0) - \omega t' \right].
\label{costheta}
\end{equation}
Here, 
\begin{equation}
  \omega = \left( p'^2_\theta + \frac{p'^2_\varphi}{\sin^2 \theta(0)}
   \right )^{1/2}.
\end{equation}   
If a kick is applied to the rotor at $t'=0$, it gains the momentum
\begin{equation}
   p'_\theta = p'_\theta(0) - P' \sin \theta(0).
\end{equation}   
Here, $p'_{\theta}(0) = p_{\theta}(0)/p_{th}$ is the initial canonical
momentum in dimensionless units, and 
$P' = \mu |\int dt {\cal E}(t)|/(I \omega_{th})$
is the dimensionless kick-strength.
If $P' \approx 1$, the energy gained due to the kick is comparable to
the initial thermal energy of the rotor.

At zero initial temperature, $p'_\varphi(0) = p'_\theta(0) = 0$.
In this case, Eq.(\ref{costheta}) is reduced to
\begin{equation}
  \theta(t') + \theta(0) - P' t' \sin \theta(0).
\end{equation}
Note, that while $P' \rightarrow \infty$ for zero temperature, the
product $P' t'$ remains finite and corresponds to the product $P \tau$
in Eq.(6).    

At finite temperature $T$, the initial distribution function of
the rotors is
\begin{equation}
 f(\theta,\varphi,p'_\theta,p'_\varphi) = \frac{1}{Q} 
 \exp \left [ - \frac{1}{2} \left( p'^2_\theta + 
 \frac{p'^2_\varphi}{\sin^2 \theta} \right) \right],
\end{equation}
with $Q$ being the partition function. Note, the $p'_\theta$ and
$p'_\varphi$ depend on the temperature. 

In the following, we present the results of Monte-Carlo simulation of the
dynamics of a thermal ensemble of kicked rotors. Figure \ref{thermal} shows
the time evolution of the angular distribution of an ensemble of molecules
with permanent dipoles subject to a kick of strength $P^{\prime }=10$ (solid
line), $P^{\prime }=5$ (dashed line), and $P^{\prime }=1$ (dotted line). In
(a), the angular distribution is plotted at focusing time $P' t' =1 $. For a
strong kick, we can still observe focusing, although the singularities in
the zero-temperature case are replaced by pronounced peaks. Moreover, the
"repulsive centrifugal force" causes a hole around $\theta =0$. 
The widths of the hole is
increasing with increasing the temperature (or decreasing $P^{\prime }$).
The angular distribution for $P' t' =3$ and $P' t' =4.5$ is shown in Fig.~%
\ref{thermal} (b) and (c), respectively. For $P^{\prime }=10$ and $P^{\prime
}=5$ the remnants of the angular rainbow moving towards the south pole are
still seen. If the initial thermal momentum of the particles is comparable
with the momentum gained through the kick ($P^{\prime }=1)$, the kick results
only in a slight perturbation of the equilibrium distribution (see Fig.~ \ref
{thermal} (a) - (c)). 

\section{Multiple kicks: Accumulative squeezing of the rotor angular
distribution.}

In the previous chapters, we investigated the wave function of a rigid rotor
kicked by a single ultra-short laser pulse. We saw that the initially
uniform angular distribution becomes much narrower in angle after the kick.
However, the degree of orientation (or alignment) that can be achieved by a
single laser kick is limited. It has been shown by us recently that the
angular distribution of an ensemble of rotors can be further squeezed by
applying a series of short laser pulses. A generic approach, the {\em %
accumulative squeezing strategy}, was proposed in \cite{rotor}. According to
this strategy, the time-dependent angular distribution of the rotors is
measured after each pulse, and the next pulse is applied at the time of the
minimal angular spread. It was shown that for 2D rotors the width of the
angular distribution is gradually decreasing with number of kicks without
any sign of saturation \cite{Lei02}. In principle, an ''unlimited'' angular
squeezing may be achieved this way.

In the following, we show how accumulative squeezing works for 3D rotors -
with permanent dipole as well as for interaction with induced polarization.
We also study analytically the asymptotic behavior of accumulative
squeezing after a large number of kicks and show, that the squeezing is
indeed unlimited in this approach. As a measure of orientation we choose the
orientation factor defined as $O(\tau )=\langle 1-\cos \theta (\tau )\rangle 
$. In a similar way, alignment is characterized by the alignment factor $%
A(\tau )=\langle 1-\cos ^{2}\theta (\tau )\rangle $.

We first provide results for angular squeezing of an ensemble of classical
rotors. Figure \ref{ac1} shows the orientation factor (alignment factor) as
the function of the number of applied short pulses. The dots represent an
ensemble with initially zero temperature, and the crosses represent two
ensembles with different finite initial temperature. In all three cases we
observe that angular squeezing is gradually improving by applying more and
more kicks, without saturation.

As the next step, we investigate analytically the squeezing process after a
large number of kicks, i.e. when the angular distribution is already very
narrow. The most of the particles are close to the pole region at this
stage, \ so that we can neglect the spherical geometry and consider a flat
model. Moreover, since the distribution is uniform in the $\varphi -$%
coordinate, we can transform the problem to dimensionless Cartesian
coordinates $x=\theta \cos \varphi $ and $y=\theta \sin \varphi $, and
consider a distribution having the same width in $x$- and $y-$directions. \
Because the interaction with the field is almost harmonic in the polar
region, we can separate the coordinates and consider independently the
behavior of the system spread in $x$ and \ \ $y$ \ coordinates.

Therefore, let us focus on the one-dimensional accumulative squeezing of a
particle moving along $x$-axis. The classical and quantum analysis of this
situation are very similar to each other, so we provide the quantum version
here. We assume that 
\begin{equation}
\overline{x}(\tau )=\overline{p}(\tau )=0.
\end{equation}
because of the symmetry. For freely moving particles, the spatial dispersion 
$\overline{x^{2}}(\tau +\Delta \tau )$ at time $\tau +\Delta \tau $ depends
on three distribution moments $\overline{x^{2}}(\tau )$, $\overline{p^{2}}%
(\tau )$, and $\overline{xp+px}(\tau )$ defined at the initial time $\tau $: 
\begin{equation}
\overline{x^{2}}(\tau +\Delta \tau )=\overline{x^{2}}(\tau )+\overline{xp+px}%
(\tau )\Delta \tau +\overline{p^{2}}\Delta \tau ^{2}.  \label{xdisp}
\end{equation}
In the similar way: 
\begin{equation}
\overline{xp+px}(\tau +\Delta \tau )=\overline{xp+px}(\tau )+2\overline{p^{2}%
}(\tau )\Delta \tau .  \label{mixedmoment}
\end{equation}
It follows immediately from Eq.(\ref{xdisp}) that $\tau _{k}$\ is the time
moment of extreme squeezing if the mixed moment $\overline{xp+px}(\tau _{k})$%
\ is zero. Assume now that the system is kicked exactly at that instant of
time (i.e., the wave function is multiplied by a factor $\exp \left(
-iPx^{2}/2\right) $). The $\overline{x^{2}}(\tau _{k})$ remains unchanged,
however the mixed moment takes a new value 
\begin{equation}
\overline{xp+px}(\tau _{k})=-2P\overline{x^{2}}(\tau _{k}).
\end{equation}
Since it is negative, the modified wave packet will reach a new minimum in
width at $\tau _{k+1}=\tau _{k}+\Delta \tau _{k}$ , where 
\begin{equation}
\Delta \tau _{k}=\frac{P\overline{x^{2}}(\tau _{k})}{\overline{p^{2}}(\tau
_{k})+P\overline{x^{2}}(\tau _{k})}.
\end{equation}
(see Eq.(\ref{mixedmoment})). There is a gain in squeezing 
\begin{equation}
\overline{x^{2}}(\tau _{k+1})=\overline{x^{2}}(\tau _{k})\left[ 1-\frac{P%
\overline{x^{2}}(\tau _{k})}{\overline{p^{2}}(\tau _{k})+P\overline{x^{2}}%
(\tau _{k})}\right] ,
\end{equation}
but the kick has increased the spread of momentum since 
\begin{equation}
\overline{p^{2}}(\tau _{k+1})=\overline{p^{2}}(\tau _{k})+P\overline{x^{2}}%
(\tau _{k}).
\end{equation}
We introduce the new notations 
\begin{eqnarray}
\overline{x^{2}}(\tau _{k}) &=&u_{k}  \nonumber \\
\frac{1}{P}\overline{p^{2}}(\tau _{k}) &=&w_{k}.
\end{eqnarray}
The quantities $u$ and $w$ satisfy the finite-difference equation 
\begin{eqnarray}
u_{k+1} &=&u_{k}-\frac{u_{k}^{2}}{u_{k}+w_{k}}  \nonumber \\
w_{k+1} &=&w_{k}+u_{k}.
\end{eqnarray}
For large $k$ they can be replaced by differential equations 
\begin{eqnarray}
\frac{du}{dk} &=&-\frac{u^{2}}{w+u}  \nonumber \\
\frac{dw}{dk} &=&u.
\end{eqnarray}
We notice that $u=x^{2}$ is monotonically decreasing with $k$, while $w$,
that is proportional to the average kinetic energy, is gradually increasing.
The above system of differential equations may be solved exactly, as it has
an integral of motion 
\begin{equation}
u^{2}+2wu=\mbox{const.}  \label{intmo1}
\end{equation}
However, in the regime of strongly developed squeezing ($u\rightarrow 0$),
the differential equations may be further simplified to 
\begin{eqnarray}
\frac{du}{dk} &=&-\frac{u^{2}}{w}  \nonumber \\
\frac{dw}{dk} &=&u.
\end{eqnarray}
Then, the integral of motion is 
\begin{equation}
wu=\mbox{const.},
\end{equation}
which is, of course, consistent with Eq.(\ref{intmo1}) in the limit $u\ll w$%
. We finally arrive to the equation 
\begin{equation}
\frac{du}{dk}=-Cu^{3},
\end{equation}
from which one immediately obtains the asymptotic behavior 
\begin{eqnarray}
\overline{x^{2}}(\tau _{k}) &\sim &u\sim \frac{1}{\sqrt{k}}  \nonumber \\
\overline{w^{2}}(\tau _{k}) &\sim &\frac{1}{u}\sim \sqrt{k}  \nonumber \\
\Delta \tau _{k} &\sim &\frac{u}{w}\sim \frac{1}{k}.
\end{eqnarray}

\section{Summary}
In this paper we concentrated on the semiclassical time evolution of 
two-dimensional and three-dimensional kicked rotational systems. 
In the 2D case we
have developed a semiclassical description of the dynamics starting from the 
exact propagator. 
In the 3D case, the problem can be greatly simplified by approximate mapping the
semiclassical motion on a sphere to the motion on a plane. Since we are 
studying wave packets in the vicinity of the poles of a sphere, 
this approximation is rather justified.

We have shown that the quantum mechanical evolution of kicked rotors can be
understood in the frame of the theory of catastrophes, a
well known subject of mathematical physics. The kicked rotational
systems experience either two
or three catastrophes with time, depending on the dimension. Similar catastrophes
have been studied in different context in various physical systems in the past. 
An interesting
property of the kicked rotor is that the catastrophes 
appear in the time domain and
can be sharpened by simply increasing the strength of the kick.   

The formation of a cusp is the first dramatic event in the dynamics of a
kicked rotor, leading to the formation of
a sharply focused rotational wave packet. In the classical limit, the cusp 
in the angular distribution exists for arbitrary small kick 
strength. In the two-dimensional case, the focused wave packet is very
accurately described in terms of Pearcey's function. 
Due to the additional azimuthal coordinate, the wave packet in the three-dimensional 
case is expressed in terms of a modified Pearcey function.

Just after the cusp, a fold singularity is created in the wave function, similar to
the rainbow formation in scattering theory. The shape of the wave packet in the
rainbow domain is explained in
terms of Airy's function or with the uniform Airy approximation. For
two-dimensional rotors, there are two rainbows in the angular distribution 
which travel in the opposite directions and
interfere with each other. In the three-dimensional case, another spectacular catastrophe 
occurs besides the rainbow: the glory effect. 
Semiclassically it is
explained in terms of classical trajectories coalescing at $\theta = 0$ and $\pi$,
and the shape of the singular wave packet
can be well described by the uniform Bessel approximation.  

In addition to zero-temperature results, we also
considered the dynamics of kicked thermal ensembles of classical 
three-dimensional rotors.
We demonstrated that thermal effects lead to holes in the focal and glory
peaks in the angular distribution.
Finally we have investigated analytically the asymptotic behavior of
the angular squeezing process in the limit of large number of kicks. We proved
that there is no saturation in this process.

Our theory for squeezing of two-dimensional kicked rotors has already been
used to optimize the squeezing of atoms in a pulsed optical
lattice [36,37].
The orientation or alignment of molecules with electromagnetic
pulses is another subject which can  benefit from our study,
the optimal conditions for molecular alignment by trains of short laser pulses
have been recently studied in [14].

\newpage

\appendix

\section{Recurrence}

In this Appendix, we want to express $\exp \left( iP\cos ^{2}\theta \right) $
in terms of spherical harmonics. We start with writing 
\begin{eqnarray}
\exp \left( iP\cos ^{2}\theta \right) &=&\exp \left[ i\left( \frac{P}{2}+%
\frac{P}{2}\cos 2\theta \right) \right]  \nonumber \\
&=&\exp \left( i\frac{P}{2}\right) \sum_{l}i^{l}j_{l}\left( \frac{P}{2}%
\right) \left( 2l+1\right) P_{l}\left( \cos 2\theta \right) .
\end{eqnarray}
We then expand $P_{l}\left( \cos 2\theta \right) $ in terms of $P_{l}\left(
\cos \theta \right) $, 
\begin{equation}
P_{l}\left( \cos 2\theta \right) =\sum_{L}d_{L,l}P_{L}\left( \cos \theta
\right) .
\end{equation}
We apply the recurrence formula for Legendre polynomials to $P_{l}\left(
\cos 2\theta \right) $ and then expand $P_{l}\left( \cos 2\theta \right) $
in terms of $P_{l}\left( \cos \theta \right) $ with $\cos 2\theta
=u=2x^{2}-1 $ and $\cos \theta =x$. 
\begin{eqnarray}
(l+1)P_{l+1}(u) &=&(2l+1)uP_{l}(u)-lP_{l-1}(u).  \nonumber \\
(l+1)\sum_{L}d_{L,l+1}P_{L}(x)
&=&(2l+1)(2x^{2}-1)\sum_{L}d_{L,l}P_{L}(x)-l\sum_{L}d_{L,l-1}P_{L}(x).
\end{eqnarray}
Then, we multiply this equation with $P_{L^{\prime }}$, integrate over $x$
and use the orthogonality relation for Legendre polynomials 
\begin{equation}
\int P_{L}(x)P_{L^{\prime }}(x)dx=\frac{2}{2L+1}\delta _{L,L^{\prime }}
\end{equation}
and arrive at 
\begin{equation}
(l+1)d_{L,l+1}\frac{2}{2L+1}=(2l+1)\left[ \left( \sum_{L^{\prime
}}d_{L^{\prime },l}\int 2x^{2}P_{L}(x)P_{L^{\prime }}(x)dx\right) -\frac{2}{%
2L+1}d_{L,l}\right] -\frac{2l}{2L+1}d_{L,l-1}.
\end{equation}
Then, we introduce the (symmetric) matrix $N_{L,L^{\prime }}=\int
P_{L}(x)P_{L^{\prime }}(x)x^{2}dx$ which has the matrix elements 
\begin{eqnarray}
N_{L,L^{\prime }}=\frac{2}{(2L+1)^{2}}\left[ \frac{(L+1)^{2}}{2L+3}+\frac{%
L^{2}}{2L-1}\right] \quad \quad &\mbox{for}&\quad \quad L^{\prime }=L 
\nonumber \\
N_{L,L^{\prime }}=\frac{2(L+1)(L+2)}{(2L+1)(2L+3)(2L+5)}\quad \quad &%
\mbox{for}&\quad \quad L^{\prime }=L+2  \nonumber \\
N_{L,L^{\prime }}=\frac{2L(L-1)}{(2L-3)(2L-1)(2L+1)}\quad \quad &\mbox{for}%
&\quad \quad L^{\prime }=L-2
\end{eqnarray}
and $N_{L,L^{\prime }}=0$ otherwise. The recurrence relation is then 
\begin{equation}
d_{L,l+1}=\frac{(2l+1)(2L+1)}{l+1}\left[ \sum_{L^{\prime }}d_{L^{\prime
},l}N_{L,L^{\prime }}\right] -\frac{2l+1}{l+1}d_{L,l}-\frac{l}{l+1}d_{L,l-1}.
\end{equation}
From the properties of $N_{L,L^{\prime }}$ follows the selection rule $%
d_{L,l}=0$ if $L$ is odd. Furthermore, since $P_{L}(1)=1$, 
\begin{equation}
\sum_{L}d_{L,l}=1.
\end{equation}
Note, however, that $d_{L,l}\neq 0$ if $l$ is odd.

\section{Asymptotics for the wave function in the focal point for the planar
model}

Considering an ensemble of classical rotors, we know that at focal time all
particles for which $\sin \theta _{0}\approx 0$ are at the (north-) pole of
the sphere. We want to investigate the semiclassical wave function in order
to see if these considerations are also valid in the semiclassical
approximation. Therefore, we calculate the wave function at focal time $%
P\tau =1$ at $\theta =0$: 
\begin{equation}
\psi \left( 0,\frac{1}{P}\right) =\frac{1}{i\sqrt{4\pi }}\exp \left[ i\left(
P+\frac{\theta ^{2}}{2\tau }\right) \right] I,
\end{equation}
with 
\begin{equation}
I=P\int_{0}^{L}d\theta _{0}\theta _{0}\exp \left[ i\frac{P}{24}\theta
_{0}^{4}\right] ,
\end{equation}
where $L$ is the radius of the circle. Expressing the exponential function
as a power series, we can perform the integration and obtain 
\begin{equation}
I=\frac{PL^{2}}{2}\sum_{\nu =0}^{\infty }\frac{\left( i\frac{P}{24}\right)
^{\nu }}{\nu !}\frac{1}{2\nu +1}L^{4\nu }.
\end{equation}
With 
\begin{equation}
\frac{1}{2\nu +1}=\frac{(2\nu )!}{2\nu +1)!}=\frac{\left( \frac{1}{2}\right)
_{\nu }}{\left( \frac{3}{2}\right) _{\nu }},
\end{equation}
we can express the sum as a confluent Hypergeometric function, 
\begin{equation}
I=\frac{PL^{2}}{2}_{1}F_{1}\left( \frac{1}{2},\frac{3}{2},i\frac{P}{24}%
L^{4}\right) .
\end{equation}
Now, we can investigate the asymptotic behavior. According to \cite{abr65},
for $|PL^{4}/24|^{2}\rightarrow \infty $, the integral is approximately 
\begin{equation}
I\approx -i\frac{\sqrt{6P}}{2}\left[ \sqrt{i\pi }+\frac{2}{L^{2}}\sqrt{\frac{%
6}{P}}\exp \left( i\frac{P}{24}L^{4}\right) \right] .
\end{equation}
The absolute square of the wave function is therefore 
\begin{equation}
\left| \psi \left( 0,\frac{1}{P}\right) \right| ^{2}\approx \frac{3P}{8\pi }%
\left[ \pi +\frac{24}{PL^{4}}+\left( \frac{24}{PL^{4}}\right) ^{1/2}\sqrt{%
\pi }\cos \left( \frac{P}{24}L^{4}+\frac{\pi }{4}\right) \right] .
\end{equation}
We can see, that for $P\rightarrow \infty $, the function goes to the
constant value $3P/8$, which is independent of the radius $L$. This is in
accordance to the classical results, where we find a singularity at $\theta
=0$.

\newpage
\centerline{FIGURE CAPTIONS}

\begin{figure}[h]
\caption{2D kicked rotor. Classical map representing the final angle $%
\protect\theta $ as a function of initial angle $\protect\theta _{0}$ for
(a) $\protect\tau =0.5\protect\tau _{f},$ (b)$\protect\tau =\protect\tau
_{f},$ (c) $\protect\tau =3\protect\tau _{f},$ and (d) $\protect\tau =10%
\protect\tau _{f}$.}
\label{clmap2}
\end{figure}


\begin{figure}[h]
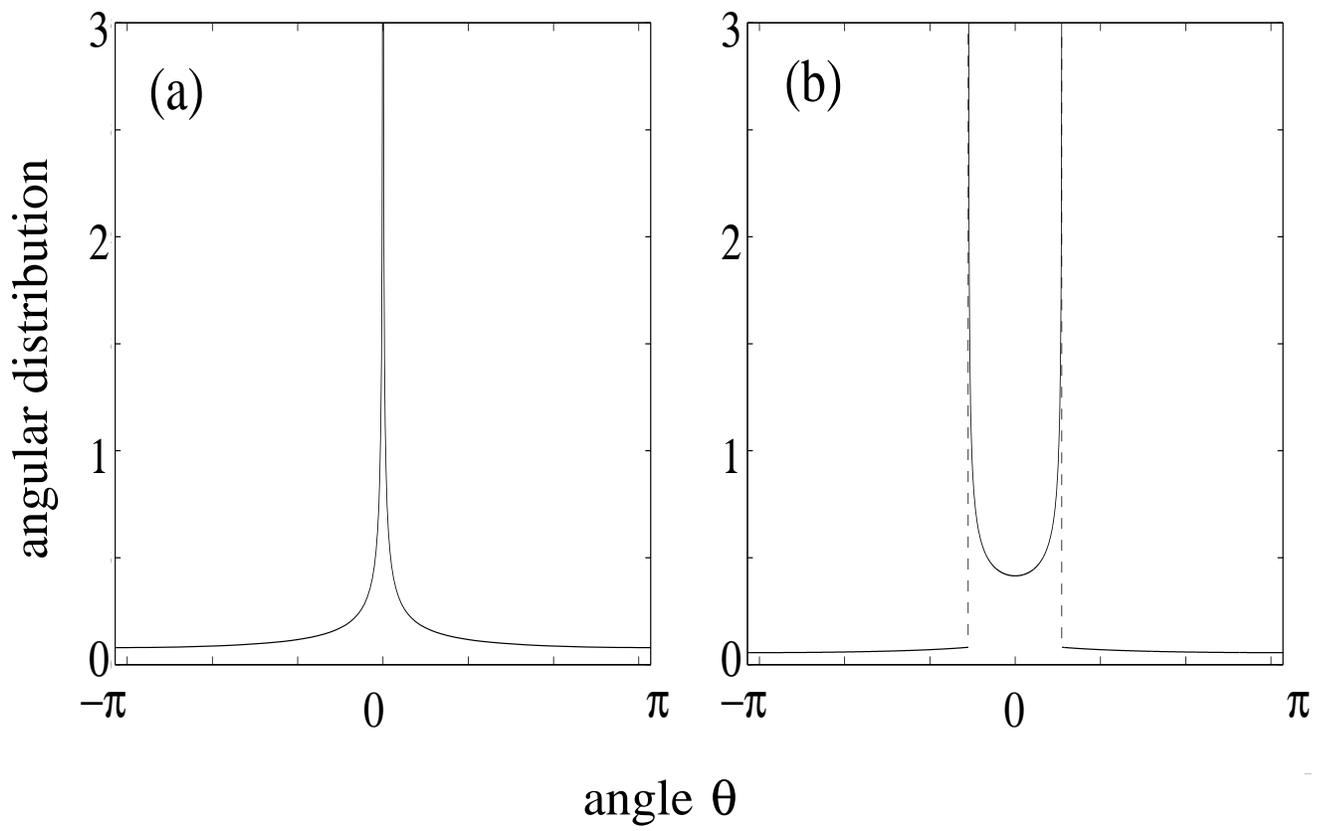

\caption{Angular distribution of the initially uniform classical ensemble 
of 2D rotors after a single $\protect\delta$-kick. Figure (a) shows focusing
at $\protect\tau=\protect\tau_f$, and (b) displays rainbows at $\protect\tau%
=1.84 \protect\tau_{f}$.}
\label{focrnb2}
\end{figure}


\begin{figure}[h]
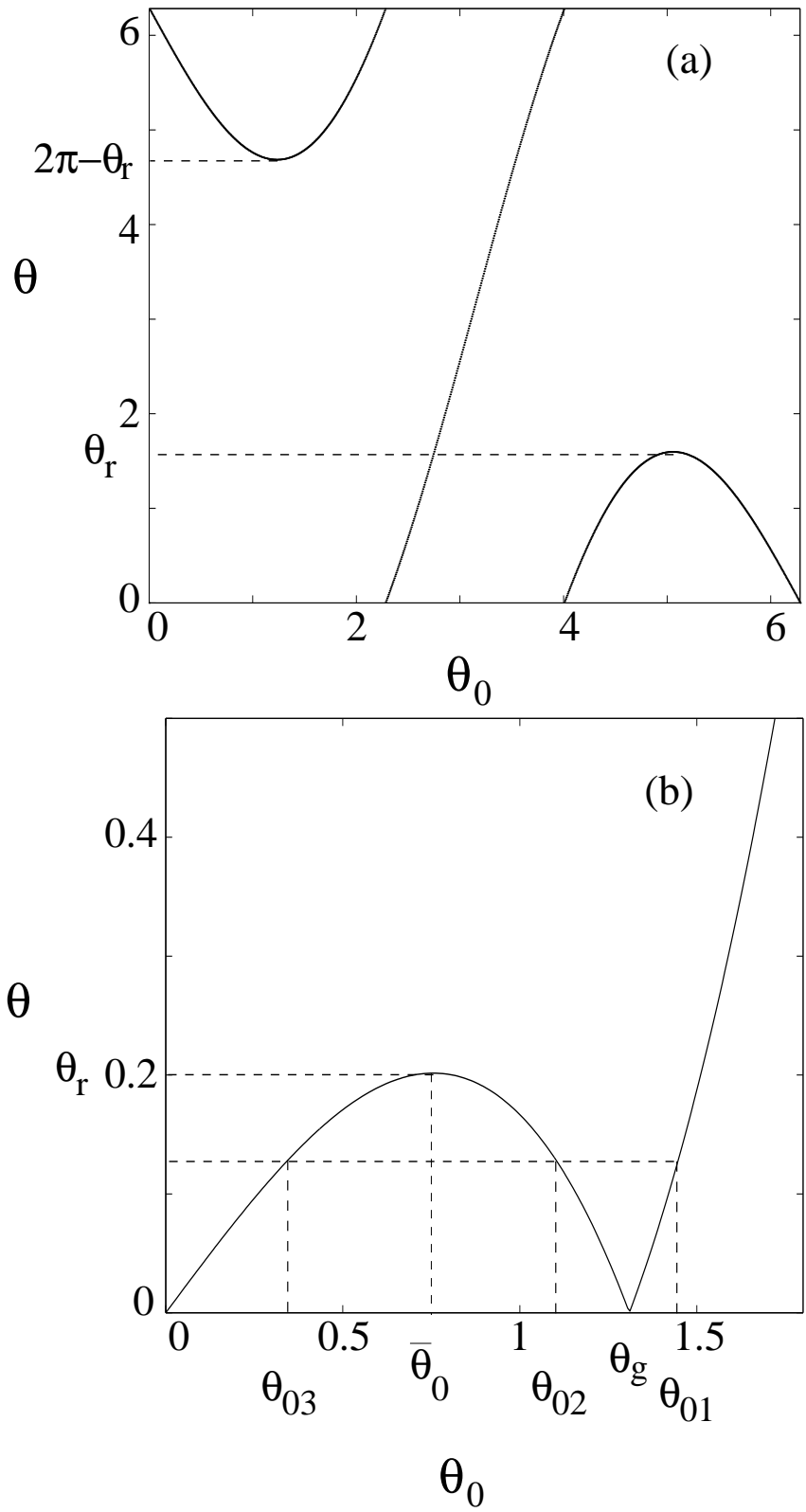

 \caption{Classical maps and critical angles for (a) 2D, and (b) 3D rotors.
In the 2D case, the final angle is plotted modulo $2 \protect\pi$ in order
to fold it into the interval $[0,2 \protect\pi]$. In (a) $P\protect\tau =3$,
and in (b) $P \protect\tau =1.4$.}
\label{clmap3}
\end{figure}


\begin{figure}[h]
\caption{Angular distribution of a linear molecule (rigid rotor) with
permanent dipole moment. In (a), (b), and (c) the probability distribution
is shown for $P \protect\tau=1$, $P \protect\tau=2$, and $P\protect\tau=4$,
respectively. In the top line, the probability of finding a classical rotor
inside the solid angle element $\sin \protect\theta d \protect\theta d 
\protect\varphi$ is plotted on a sphere. The middle line shows the classical
distribution function as a function of $\protect\theta$, and the bottom line
displays the corresponding quantum mechanical probability distribution for a
kick-strength of $P=75$.}
\label{orient}
\end{figure}


\begin{figure}[h]
\caption{Angular distribution of a linear molecule (rigid rotor) interacting
with the external field due to induced polarization. In (a), (b), and (c)
the probability distribution is shown for $P\protect\tau =0.5$, $P\protect\tau
=2$, and $P\protect\tau =2.4$, respectively. In the top line, the
probability of finding a classical rotor inside the solid angle element $%
\sin \protect\theta d\protect\theta d\protect\varphi $ is plotted on a
sphere. The middle line shows the classical distribution function as a
function of $\protect\theta $, and the bottom line displays the
corresponding quantum mechanical probability distribution for a
kick-strength of $P=75$.}
\label{align}
\end{figure}


\begin{figure}[h]
\caption{Angular distribution of a 2D quantum rotor excited by a strong $%
\protect\delta $-kick ($P=85$). The graphs correspond to (a) $\protect\tau %
=0.5\protect\tau _{f},$ (b)$\ \protect\tau =\protect\tau _{f},$ (c) $\protect%
\tau =2\protect\tau _{f}$, (d) $\protect\tau =\protect\tau _{f}+T_{rev}/2,$
(e) $\protect\tau =\protect\tau _{f}+T_{rev}/3,$ (f) $\protect\tau =\protect%
\tau _{f}+T_{rev}/4,$ (g) $\protect\tau =1.8\protect\tau _{f}+T_{rev}/2,$
(h) $\protect\tau =1.8\protect\tau _{f}+T_{rev}/3,$ and (i) $\protect\tau %
=1.8\protect\tau _{f}+T_{rev}/4,$ respectively.}
\label{2Dq}
\end{figure}


\begin{figure}[h]
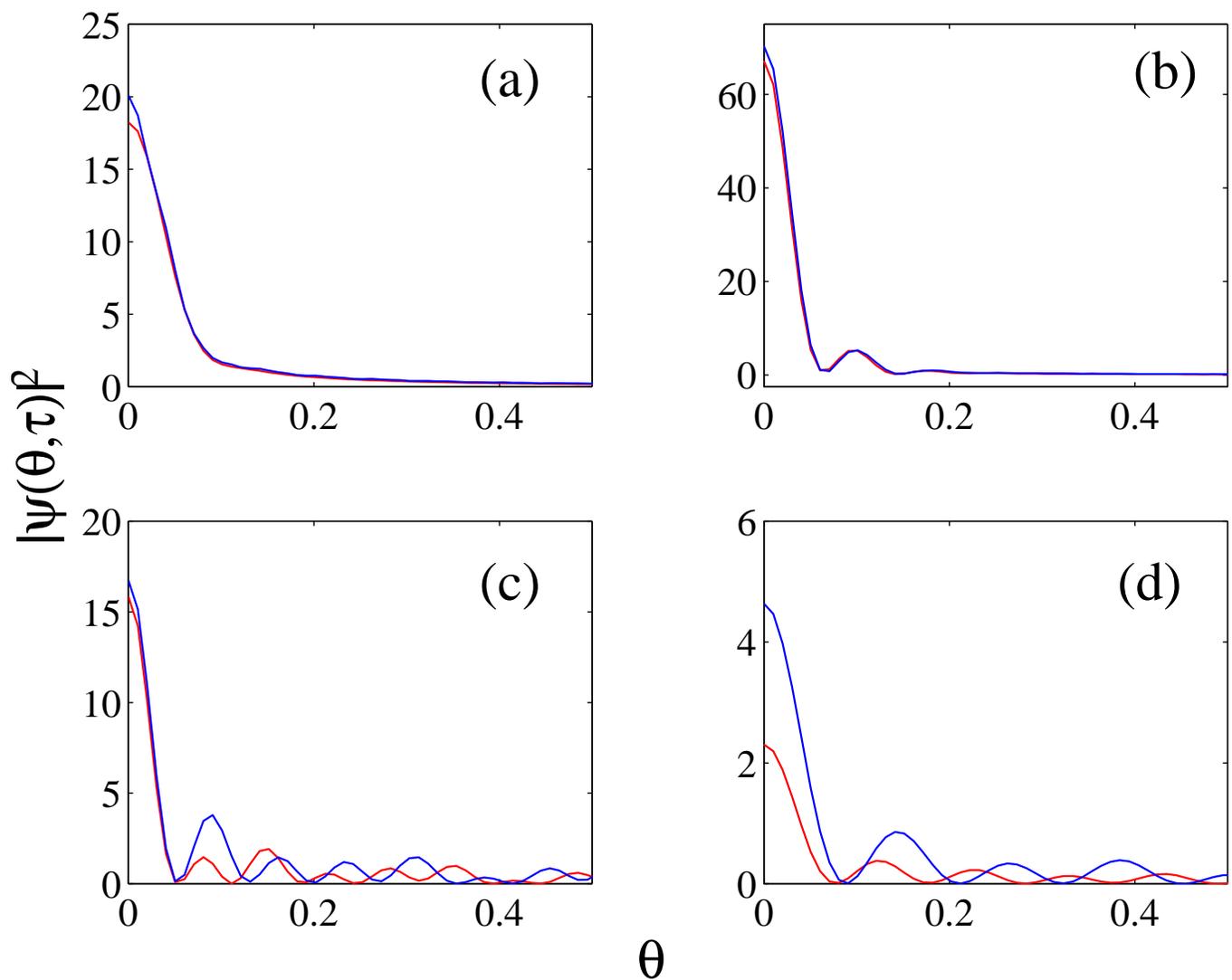

\caption{Probability distribution in $\protect\theta$ for the 3D kicked
rotor (solid line) and  probability distribution for the planar model
(dashed line). The figure shows the probability distribution
for $\protect\tau=1/P$ (a), 
$\protect\tau=1.2/P$ (b), $\protect\tau=2/P$ (c), and $\protect\tau=4/P$ (d).
The kick strength is $P=50$.}
\label{pmod}
\end{figure}


\begin{figure}[ht]
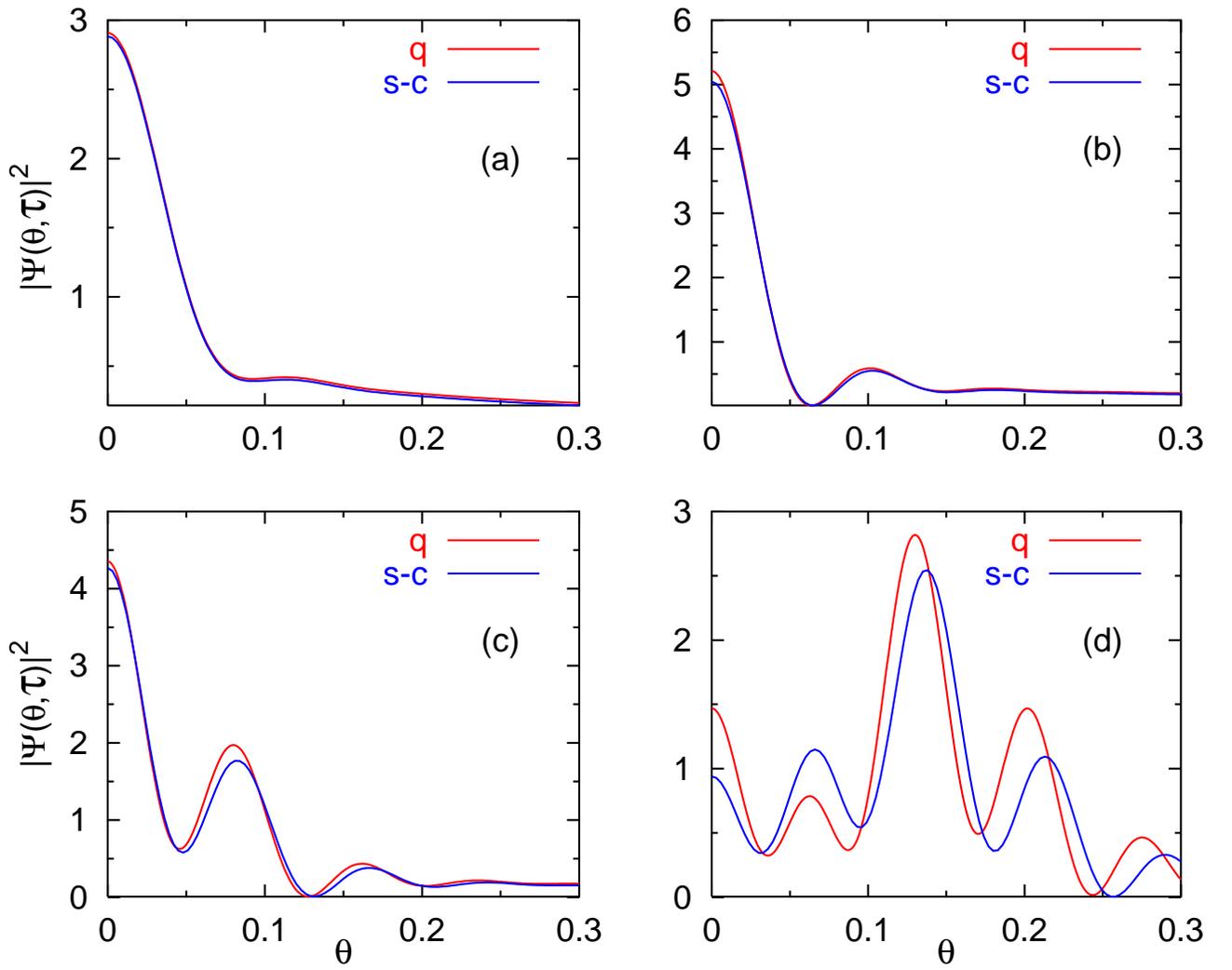

\caption{Figure (a) shows the probability distribution (2D case)
at focal time $\protect\tau=1/P$.
In (b), (c), and (d), $\protect\tau=1.1/P$, $\protect\tau=1.2/P$, and $%
\protect\tau=1.4/P$, respectively. The kick strength is $P=50$. The solid line
represents the exact quantum mechanical probability distribution, while the
dashed line described the probability distribution in Pearcey approximation.}
\label{2dim7}
\end{figure}


\begin{figure}[ht]
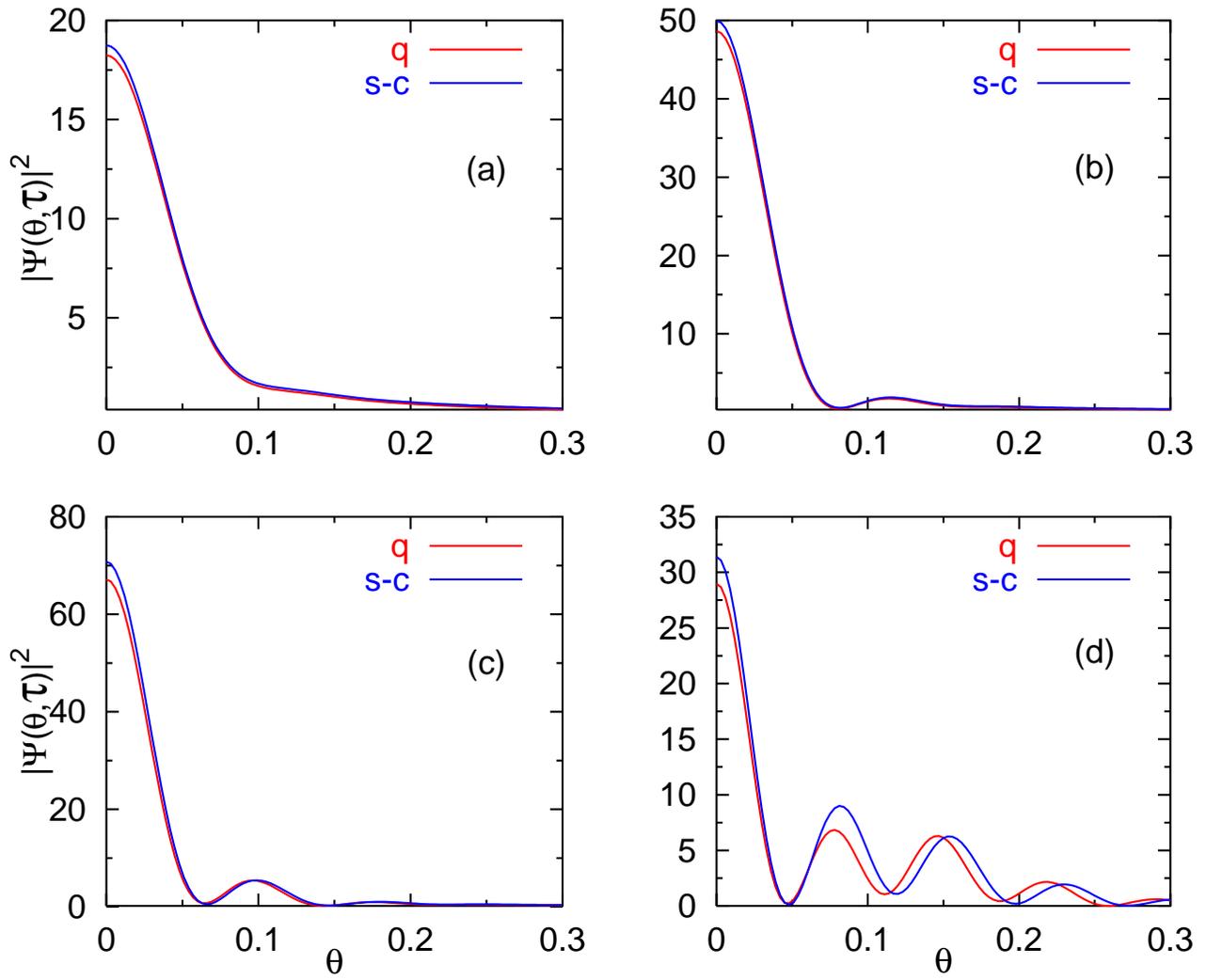

\caption{Glory and oscillations close to the focusing in the kicked 3D
rotor. Figure (a) shows the probability distribution at focal time $\protect\tau=1/P$.
In (b), (c), and (d), $\protect\tau=1.1/P$, $\protect\tau=1.2/P$, and $%
\protect\tau=1.4/P$, respectively. The kick strength is $P=50$. The solid line
represents the exact quantum mechanical probability distribution, while the
dashed line shows the probability distribution obtained in the approximation
given by Eq.~(\ref{cusp1}) based on Pearcey function.}
\label{3dim7}
\end{figure}


\begin{figure}[thb]
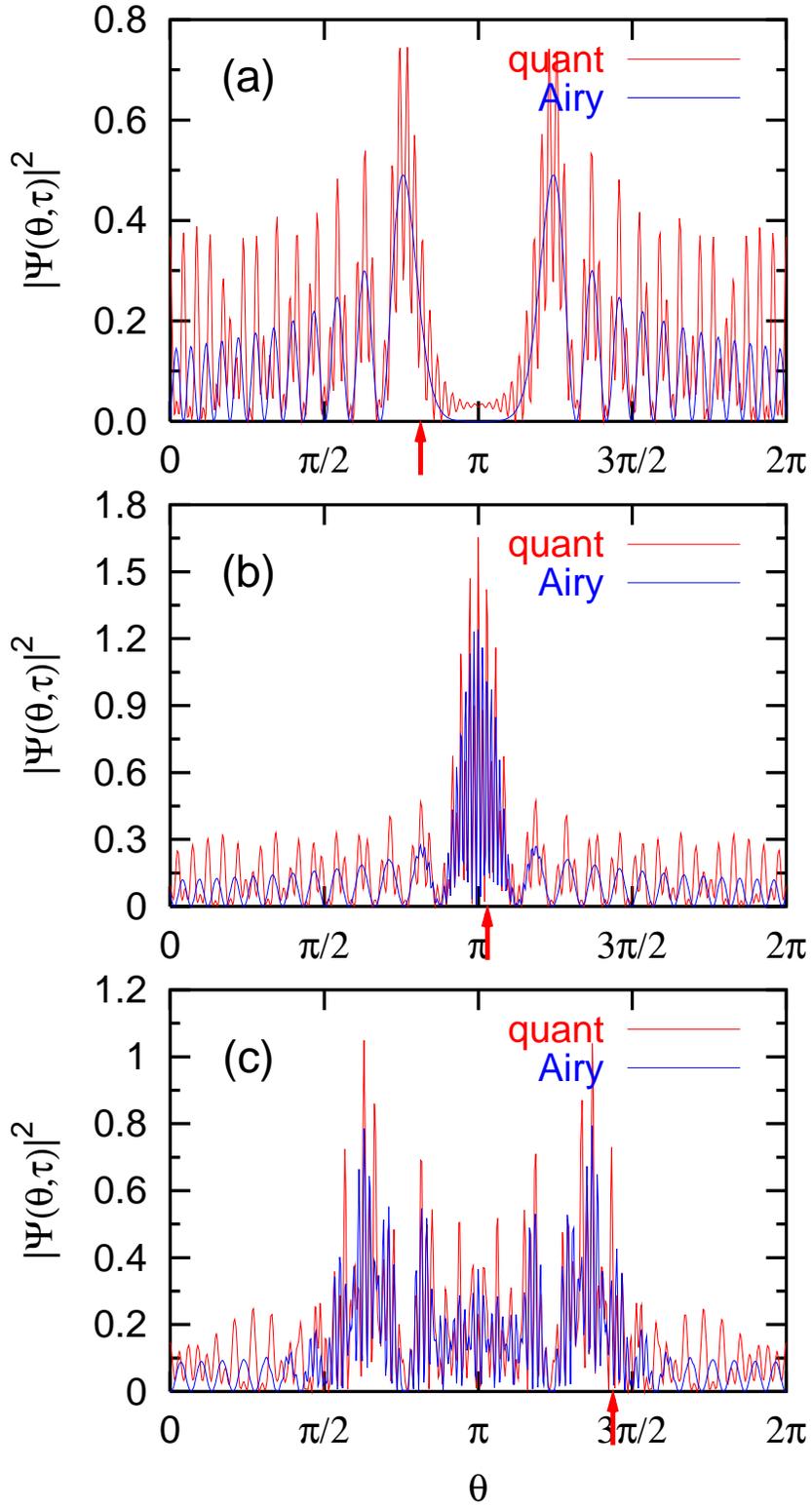

\caption{Airy approximation for 2D WP evolution for times (a) $\protect\tau %
P $=4, (b) 4.7 and (c) 6, $P$=75. The solid line represents exact quantum
result, the dashed one shows the Airy approximation. The arrow point to the
rainbow angle $\protect\theta_{r}$. }
\label{ffairy}
\end{figure}


\begin{figure}[tbh]
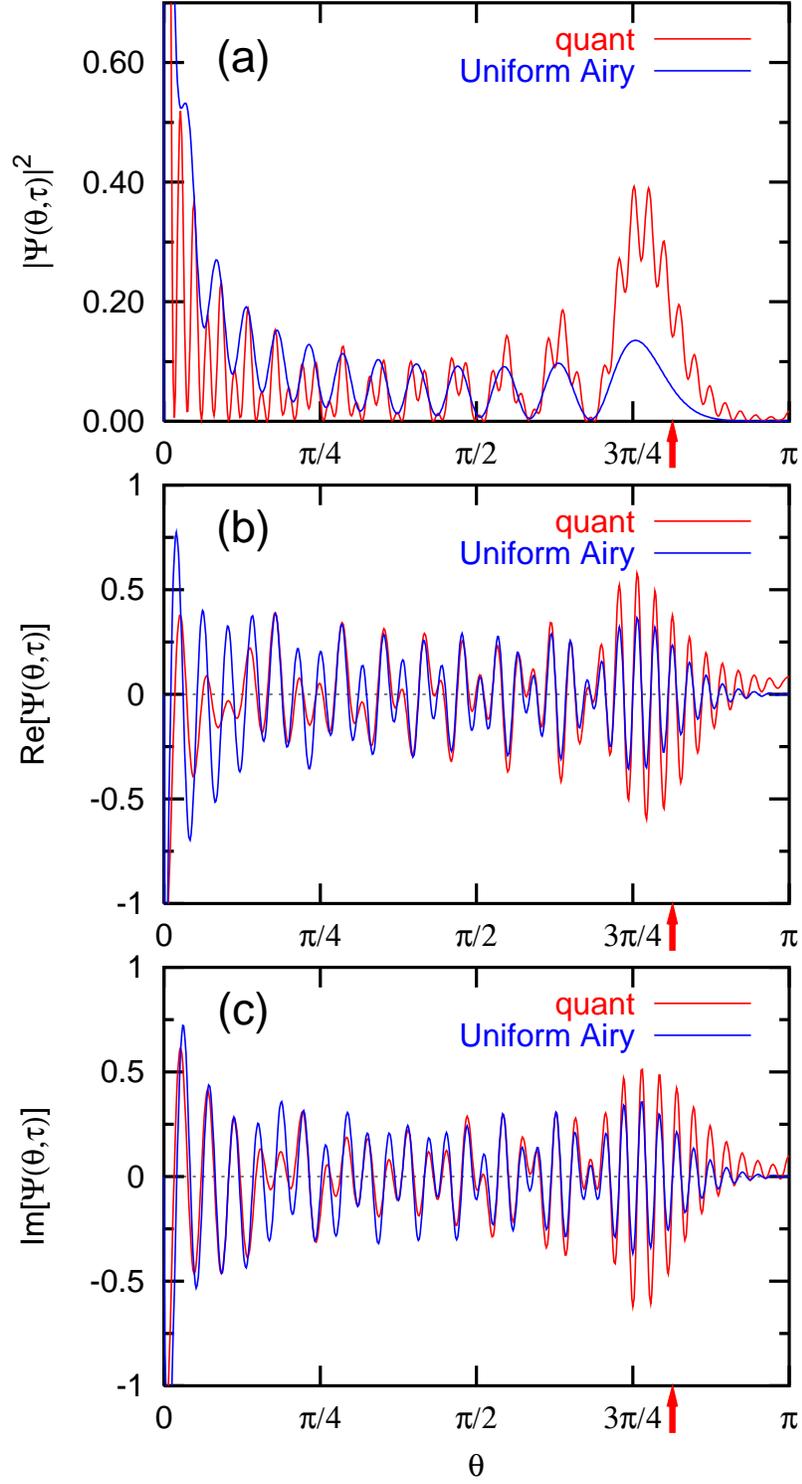

\caption{Uniform Airy approximation in the 3D case (Eq.~(\ref{e17})) is
compared to quantum result for $P\protect\tau $=4 and $P$=75. The arrow
points to the rainbow angle. Figure (a) shows the probability density as a
function of $\protect\theta $, and figures (b) and (c) show the real and the
imaginary part of the wavefunction.}
\label{f8}
\end{figure}


\begin{figure}[h]
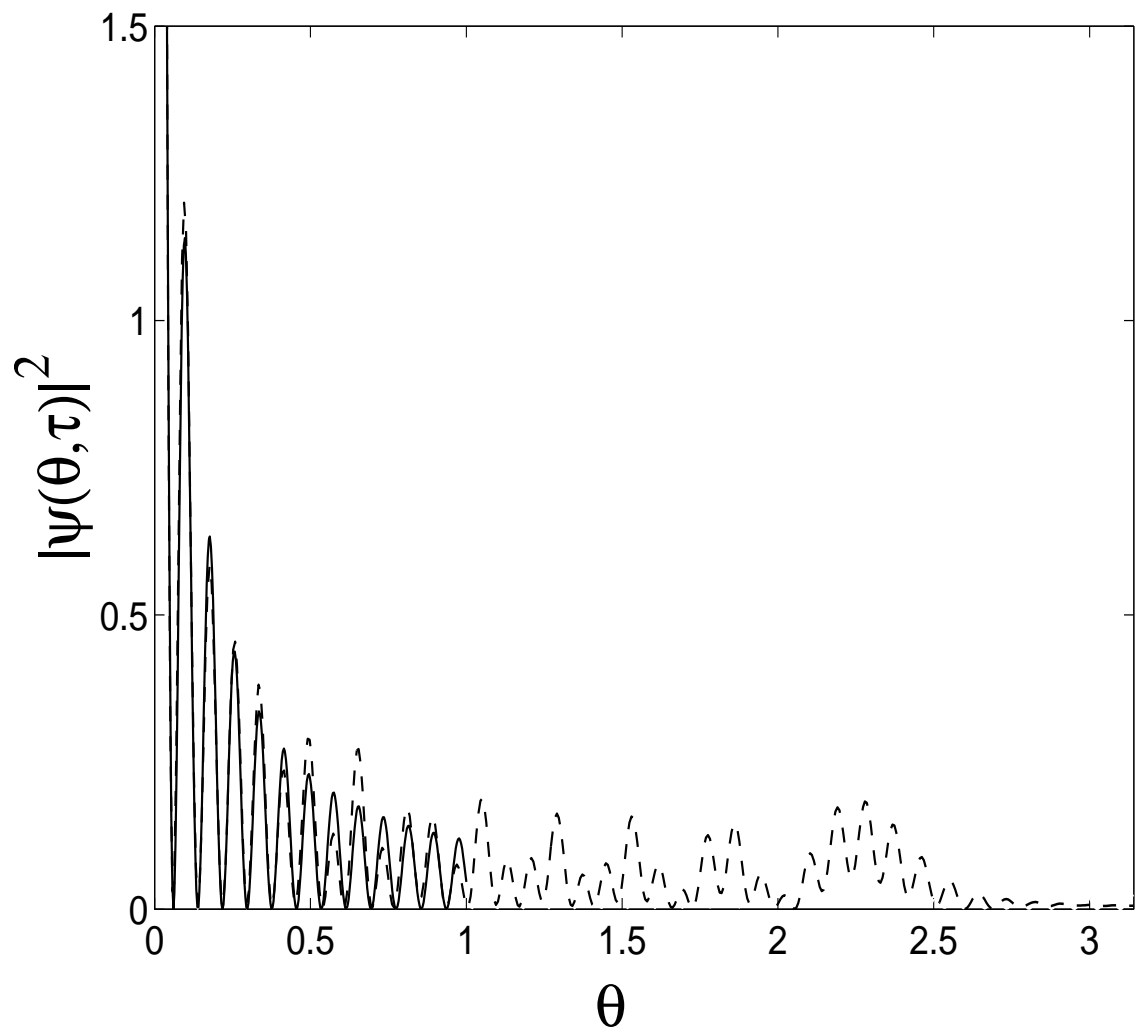

\caption{Absolute square of the wave function in uniform Bessel
approximation (solid line) compared to the exact wave function (dashed line)
for $P \protect\tau =4$ and $P=75$ (3D case).}
\label{bessel_fig}
\end{figure}

\begin{figure}[h]
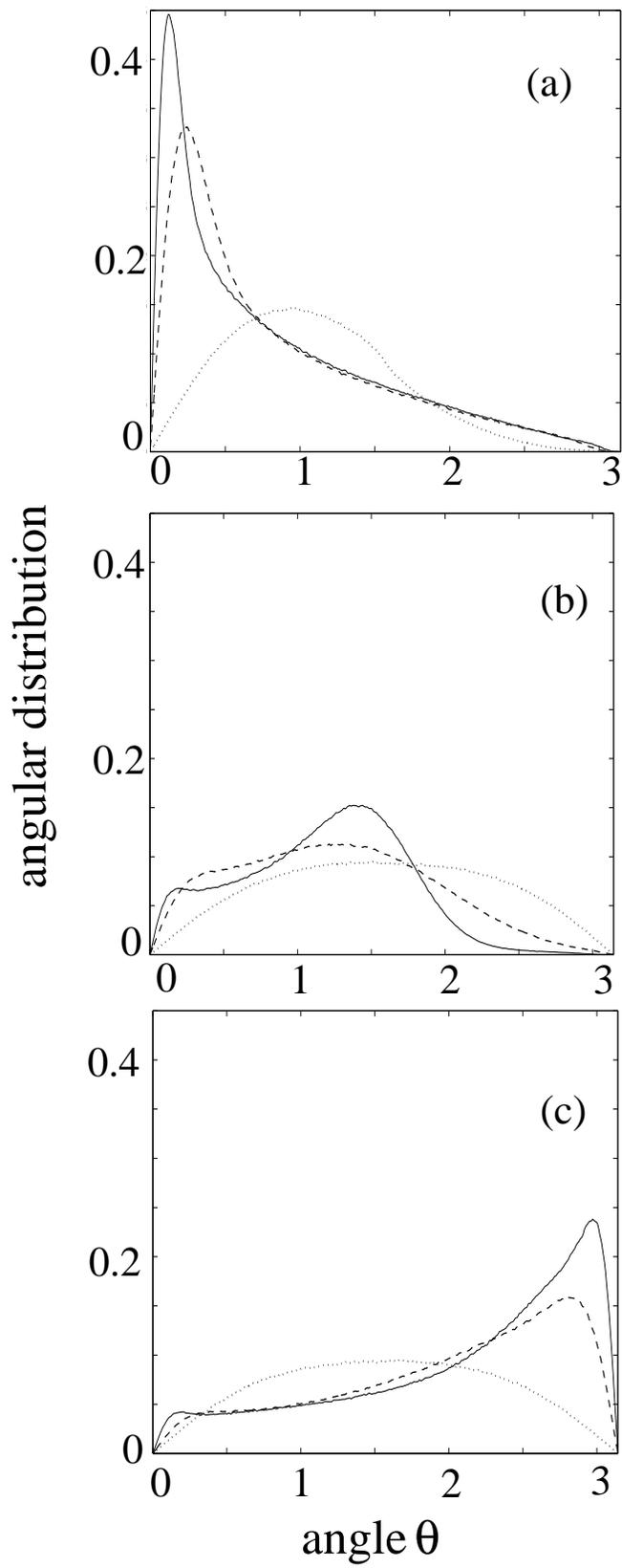

\caption{Angular distribution of an ensemble of classical rotors (in 3D)
with finite
initial temperature for $P' \protect t' =1$ (a), $P' \protect t' =3$ (b),
and $P' \protect t' = 4.5$ (c). The solid line corresponds to the kick
strength $P^{\prime}=10$, the dashed line to $P^{\prime}=5$ and the dotted
line to $P^{\prime}=1$. }
\label{thermal}
\end{figure}


\begin{figure}[h]
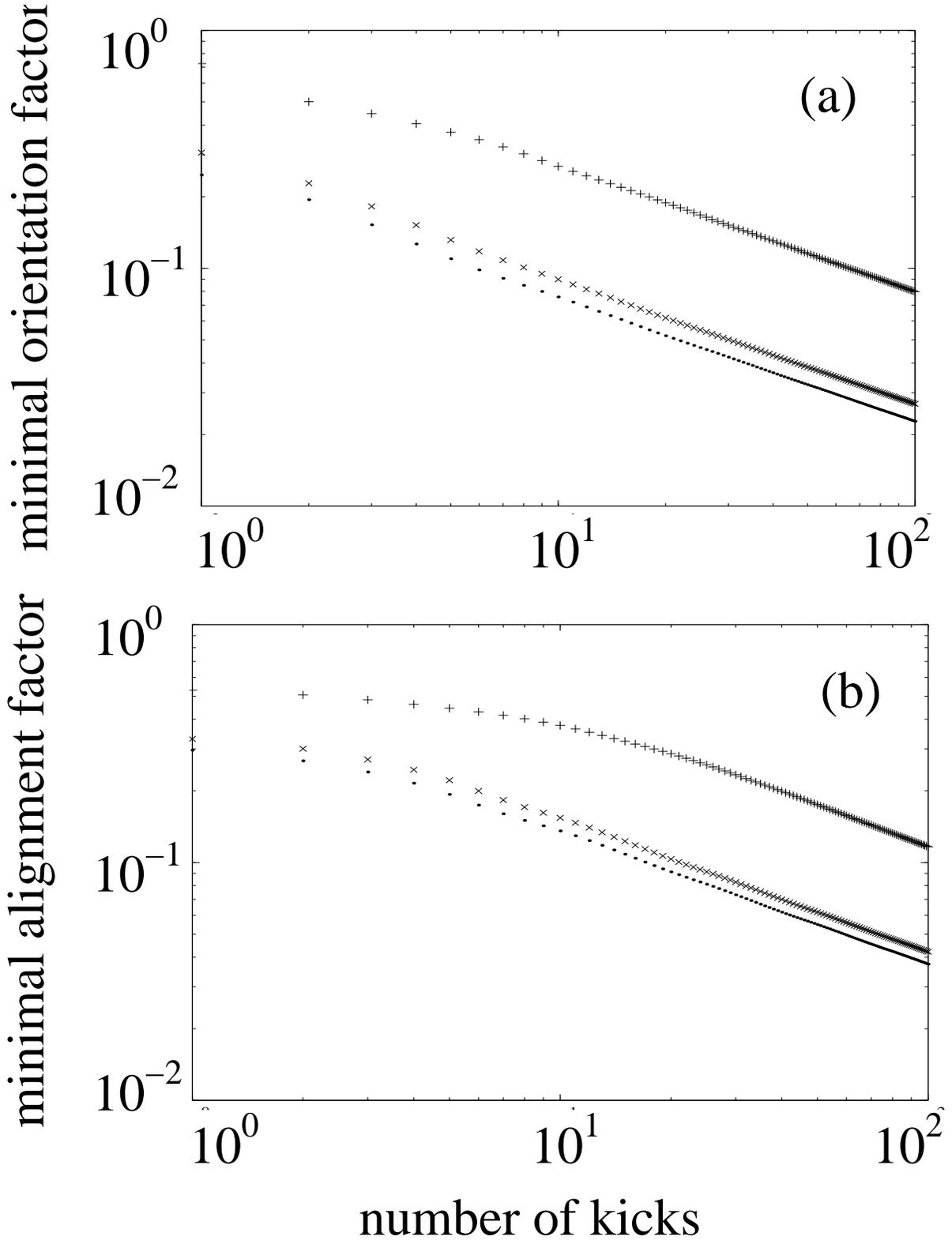

\caption{Accumulative squeezing of a classical ensemble of 3D rotors.  
Figure (a) shows the minimal orientation factor, and
figure (b) shows the minimal alignment factor as a function of the number of
kicks. The dots (.) correspond to an ensemble of rotors with zero initial
temperature ($P^{\prime }=\infty $). The crosses (X) and (+) correspond to $%
P^{\prime }=5$ and $P^{\prime }=1$, respectively. The orientation factor and
alignment factor were calculated with help of Monte Carlo simulation with
10000 particles}
\label{ac1}
\end{figure}

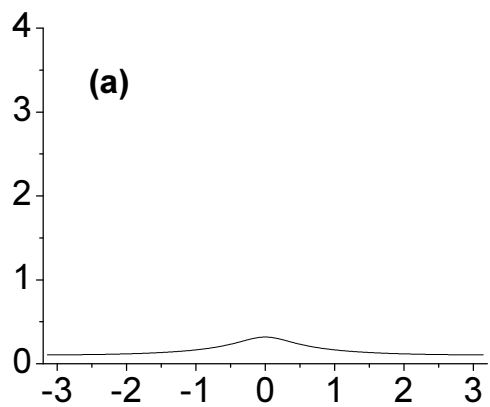
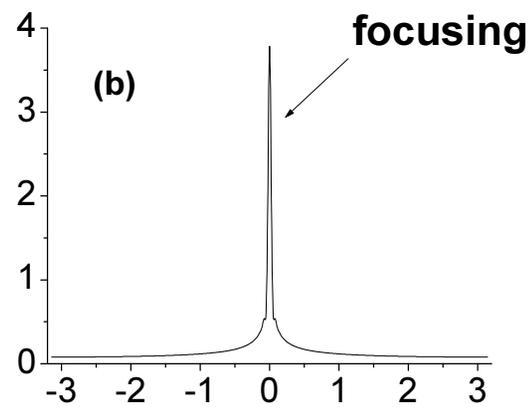
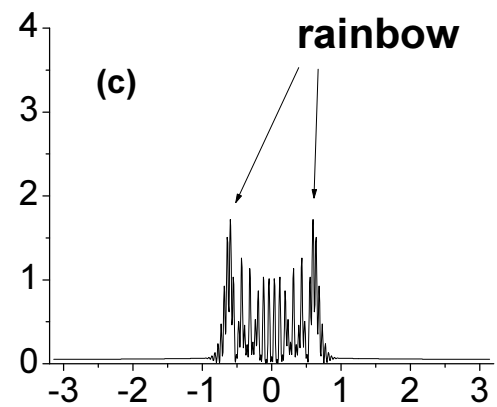
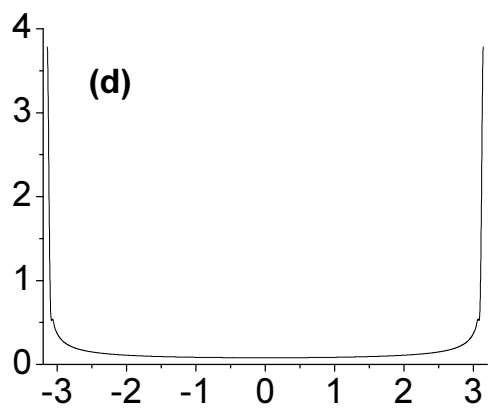
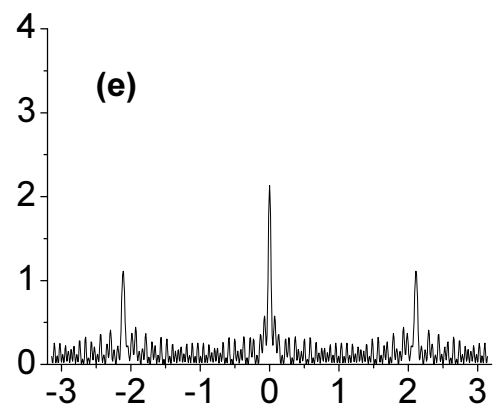
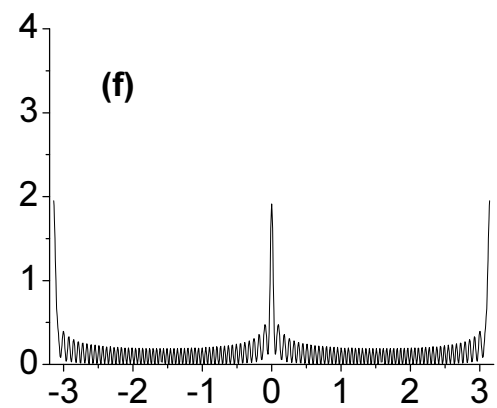
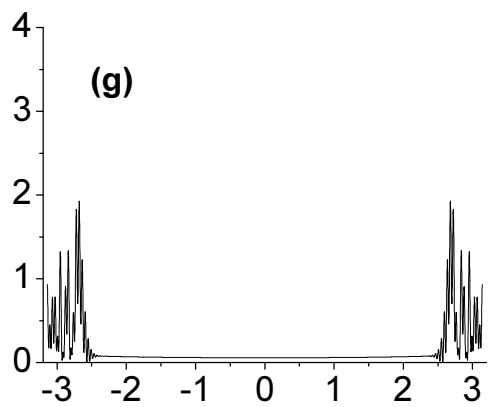
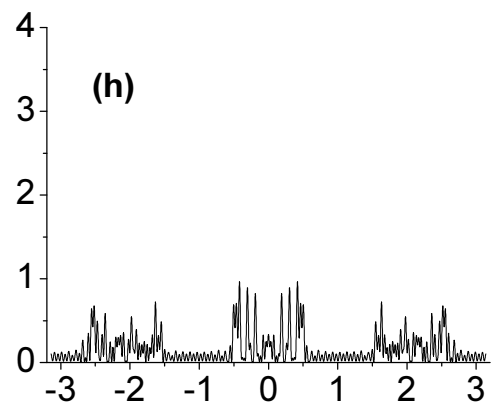
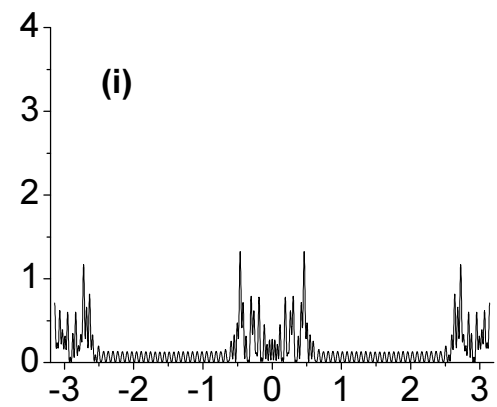
\end{document}